\documentclass[preprint,prd,aps,a4paper,10pt,titlepage,superscriptaddress]{revtex4-1}
\usepackage{ucs}
\usepackage[utf8x]{inputenc}
\usepackage{amsmath}
\usepackage{amsfonts}
\usepackage{amssymb}
\usepackage{amsthm}
\usepackage[english]{babel}
\usepackage[T1]{fontenc}
\usepackage{graphicx}
\usepackage[dvips]{hyperref}

\begin{document}
\title{Theoretical framework to analyze searches for hidden light gauge bosons in electron scattering fixed target experiments
}
\author{T. Beranek}
\author{H. Merkel}
\author{M. Vanderhaeghen}
\affiliation{PRISMA Cluster of Excellence, Johannes Gutenberg-Universit\"at Mainz, D-55099 Mainz}
\affiliation{Institut f\"ur Kernphysik, Johannes Gutenberg-Universit\"at Mainz, D-55099 Mainz}
\date{\today}

\newcommand{\fslash}[1]{(\gamma\cdot{#1})}
\newcommand{\Abs}[1]{|\vec{#1}\,|}
\newcommand{\Absp}[1]{|\vec{#1}^{\,\prime}|}
\newcommand{\vecp}[1]{\vec{#1}^{\,\prime}}
\newcommand{\ubar}{{\overline{u}}}
\newcommand{\vbar}{{\overline{v}}}
\newcommand{\Ap}{{\gamma^\prime}}
\newcommand{\MeV}{\mathrm{MeV}}
\newcommand{\GeV}{\mathrm{GeV}}
\newcommand{\D}{\mathrm{D}}
\newcommand{\X}{\mathrm{X}}

\begin{abstract}
Motivated by anomalies in cosmic ray observations and by attempts to solve questions of the Standad Model of particle physics like the $(g-2)_\mu$ discrepancy, $U(1)$ extensions of the Standard Model have been proposed in recent years.
Such $U(1)$ extensions allow for the interaction of Dark Matter by exchange of a photon-like massive force carrier $\Ap$ not included in the Standard Model.
In order to search for $\Ap$ bosons various experimental programs have been started.
One approach is the dedicated search at fixed-target experiments at modest energies as performed at MAMI or at the Jefferson Lab.
In these experiments the process $e(A,Z)\rightarrow e(A,Z)l^+l^-$ is investigated and a search for a very narrow resonance in the invariant mass distribution of the $l^+l^-$ pair is performed.
In this work we analyze this process in terms of signal and background in order to describe existing data obtained by the A1 experiment at MAMI with the aim to give accurate predictions for exclusion limits in the $\Ap$ parameter space.
We present a detailed theoretical analysis of the cross sections entering in the description of such processes.
\end{abstract}

\maketitle

\section{Introduction}
With the recent observation of a new boson at the LHC, which is expected to be the Higgs boson, the last missing element of the Standard Model of particle physics (SM) seems to be discovered \cite{:2012gk,:2012gu}.
Despite of this, nowadays the existence of dark matter, which is not included in the SM, is established as a necessary ingredient in order to explain the energy density of the universe within the cosmological standard model \cite{Bertone:2004pz,Komatsu:2010fb,Beringer:1900zz}.
The nature of dark matter is however still a wide open question.
Neither is it known what dark matter is made of, nor in which way it is interacting with other particles, e.g. the SM particles.
Besides the unsolved problem of dark matter, the SM itself does contain several issues like the discrepancy in the theoretical and experimental determination of the anomalous magnetic moment of the muon $(g-2)_\mu$, the proton radius puzzle, or weak scale questions like the hierarchy problem, which all could be hints for physics beyond the SM.\\
Recent observations of anomalies in astrophysical data \cite{Strong:2005zx,Adriani:2008zr,Cholis:2008wq} have motivated to consider extensions of the SM by including an additional $U(1)$ gauge group which could explain such anomalies \cite{ArkaniHamed:2008qn,Pospelov:2008jd}.
Though the idea to extend the SM by an additional $U(1)$ recently became popular, it did not rise up with the observations.
In many well motivated SM extensions, e.g. from string theory, additional $U(1)$ groups appear naturally \cite{Weinberg:1977ma,Okun:1982xi,Holdom:1986eq,Fayet:1990wx,Pospelov:2008zw,Andreas:2011in}.\\
Extending the SM by such an $U(1)_D$ group generates an additional gauge boson $\Ap$ which is able to interact with the electromagnetic current of the Standard Model.
Although this interaction is forbidden at tree level it is possible via kinetic mixing \cite{Holdom:1986eq} giving rise to an effective interaction Lagrangian
\[
 \mathcal{L}_{\text{int}}=i\,\varepsilon e\, \bar{\psi}_\text{SM}\, \gamma^\mu\, \psi_\text{SM} \,A^\prime_\mu,
\]
where $A^\prime$ denotes the $\Ap$ field.
Furthermore, $\varepsilon$ is the kinetic mixing factor parameterizing the coupling strength relative to the electric charge $e$, and describes the interaction of the additional gauge boson with the electromagnetic current.
The $\Ap$ may gain a mass $m_\Ap$ which can be estimated to be in the range of $10\, \MeV$ to a few $\GeV$ \cite{Fayet:2007ua,Cheung:2009qd,Essig:2009nc}.
The kinetic mixing factor $\varepsilon^2={\alpha^\prime}/{\alpha}$ is predicted from various models to be in the range $10^{-12} < \varepsilon < 10^{-2}$ \cite{Essig:2009nc,Goodsell:2009xc}.
Due to the coupling via kinetic mixing the $\Ap$ may decay to dark matter particles as well as SM matter particles.
In the case, that the decay to dark matter is kinematically forbidden and $m_\Ap>2m_e=1.022\,\MeV$, which this work will focus on, the $\Ap$ will decay to SM particles and therefore must be observable at accelerator experiments.\\
The $\Ap$ interacts with SM particles and has properties which are very similar to that of the photon.
Since by now such a boson could not be observed one often refers to the $\Ap$ (which is also denoted as $A^\prime$, $U$, $\phi$) as heavy, hidden, para- or dark photon.
Within this minimal model the free parameters are the mass $m_\Ap$ and the coupling strength $\varepsilon$.
In pioneering works several constraints from existing data were obtained on these parameters e.g. in beam dump searches or by the BaBar experiment, as well as from $(g-2)$ analyses \cite{Pospelov:2008zw,Bjorken:2009mm}.\\
The coupling of the $\Ap$ to SM particles and the predicted mass range allows for the $\Ap$ search by accelerator experiments at modest energies with high intensities.
While collider experiments are ideally suited for higher $\Ap$ masses, fixed-target experiments with their high luminosities are ideally suited for the $\Ap$ search in the MeV to $1$ GeV range \cite{Bjorken:2009mm,Batell:2009yf,Batell:2009di,Essig:2009nc,Reece:2009un}.
The proposal to search for the hidden gauge boson by fixed-target experiments motivated several experimental programs, both by the A1 collaboration at the MAMI accelerator in Mainz \cite{Merkel:2011ze} as well as at the CEBAF facility at Jefferson Lab with the APEX \cite{Essig:2010xa,Abrahamyan:2011gv}, HPS \cite{HPS} and DarkLight \cite{Freytsis:2009bh,Kahn:2012br} experiments.
The A1 and APEX experiments already have published first data.
Furthermore in many recent publications, constraints on the $\Ap$ parameter space from the analysis of beam dump searches \cite{Blumlein:2011mv,Gninenko:2011uv,Gninenko:2012eq,Andreas:2012mt}, meson decays and collider experiments \cite{Aubert:2009cp,Archilli:2011zc,Babusci:2012cr,Echenard:2012iq} as well as from other arguments were given \cite{Davoudiasl:2012ig,Endo:2012hp}, and are summarized in Fig.~\ref{fig:excl_compilation}.
In addition, many other experiments were proposed to probe the light hidden sector or are underway, for a review see e.g. Ref.~\cite{Hewett:2012ns}.\\
In all considered fixed-target experiments an electron beam is scattered off a fixed target which is either a proton or a heavy nucleus like tantalum.
Induced by this electromagnetic process a $\Ap$ may be radiated from the electron beam and decays into SM particles like an electron-positron pair.
Detecting the decay particles and reconstructing the invariant mass of the pair allows to search for the hidden gauge boson by a bump hunt.
The $\Ap$ will manifest itself by a very sharp peak above the radiative background that results from the corresponding process where a virtual photon is radiated from the electron beam which creates a lepton pair, too, i.e. the underlying process
\[
 e(A,Z)\rightarrow e(A,Z)l^+l^-
\]
is investigated.\\
If there is no bump seen in the invariant mass spectrum this allows to exclude regions of the $\Ap$ parameter space given by the kinetic mixing factor $\varepsilon$ and its mass $m_{\Ap}$.
In order to perform this study a precise knowledge of the signal and background cross sections are crucial.
Such precise study is the main subject of the present work.\\
This work is structured as follows: In section~\ref{sec:calculations} we present our calculations of the signal and background cross sections.
In section~\ref{sec:MAMI_results} we present our results of the cross section calculations for the experiments performed at MAMI.
Furthermore we present a comparison with available data.
In section~\ref{sec:future} we propose new searches at MAMI and at the new MESA accelerator and present our predictions for the exclusion limits.
\section{Calculation of the signal and background cross sections\label{sec:calculations}}
\begin{figure}
\includegraphics[width=.48\linewidth]{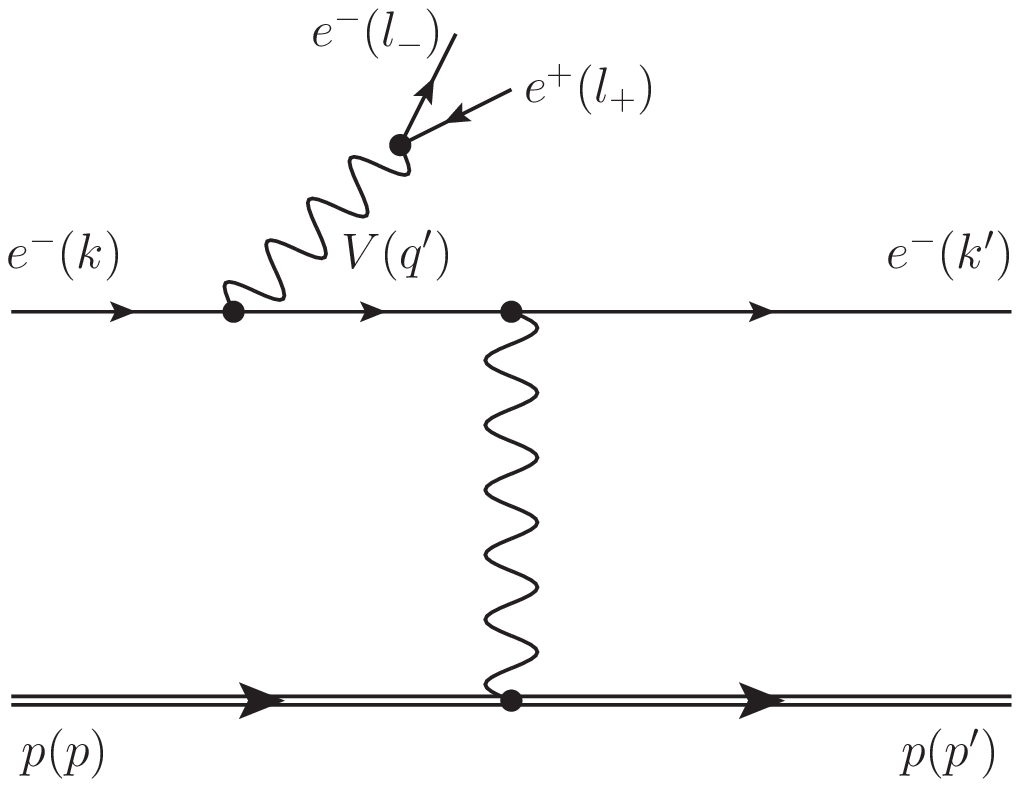} \includegraphics[width=.48\linewidth]{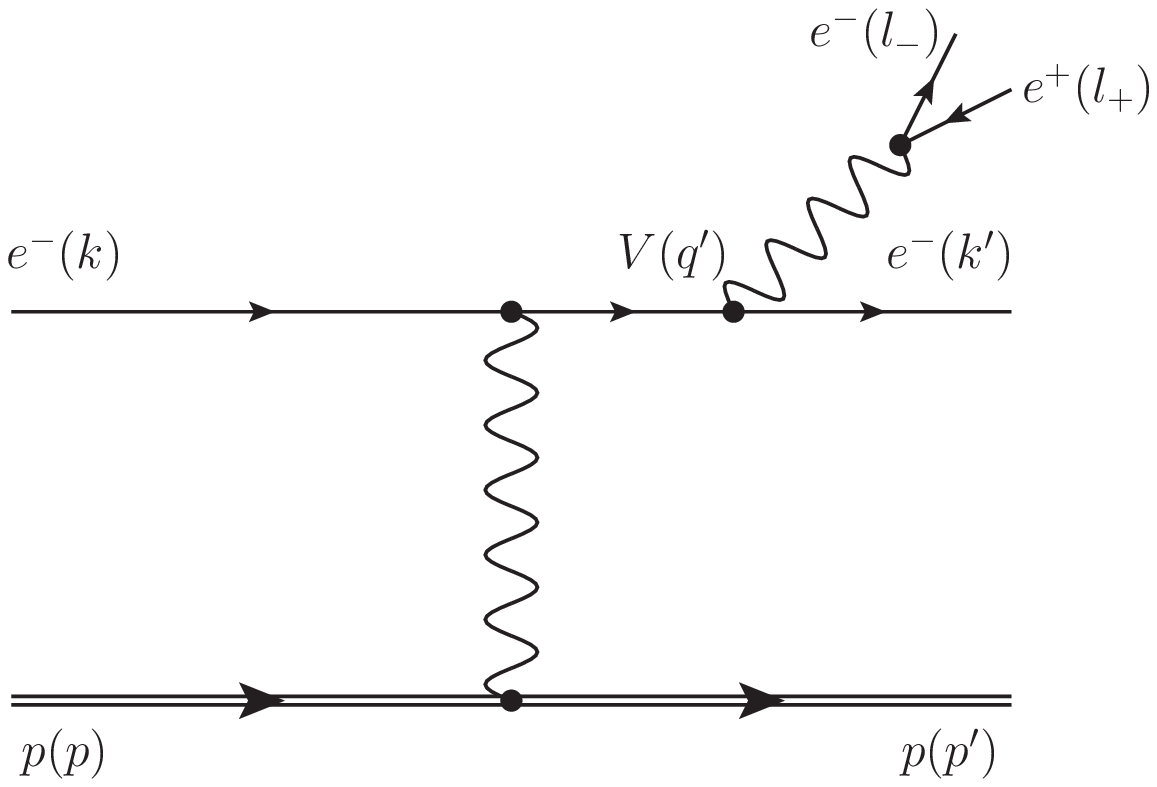}\\
\includegraphics[width=.48\linewidth]{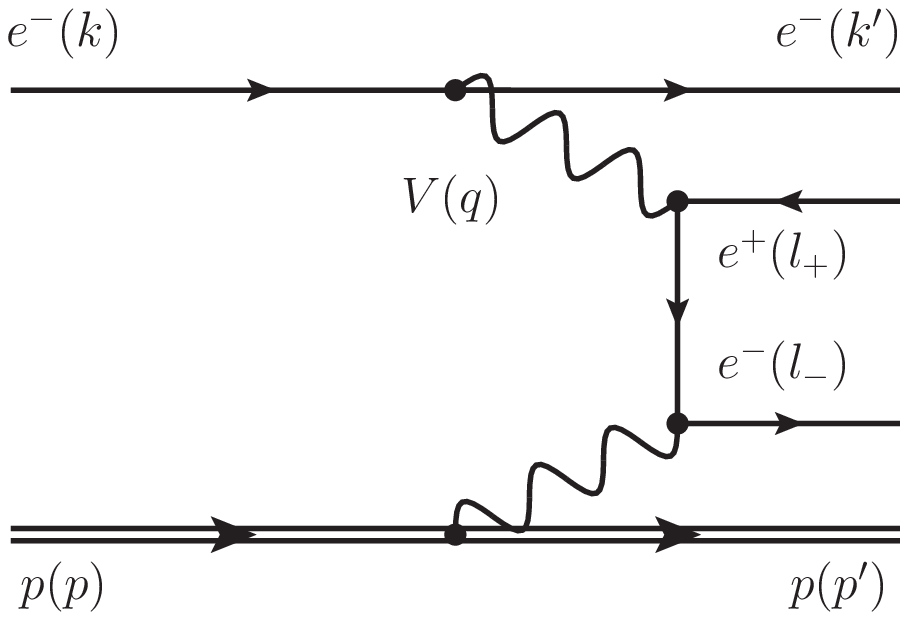} \includegraphics[width=.48\linewidth]{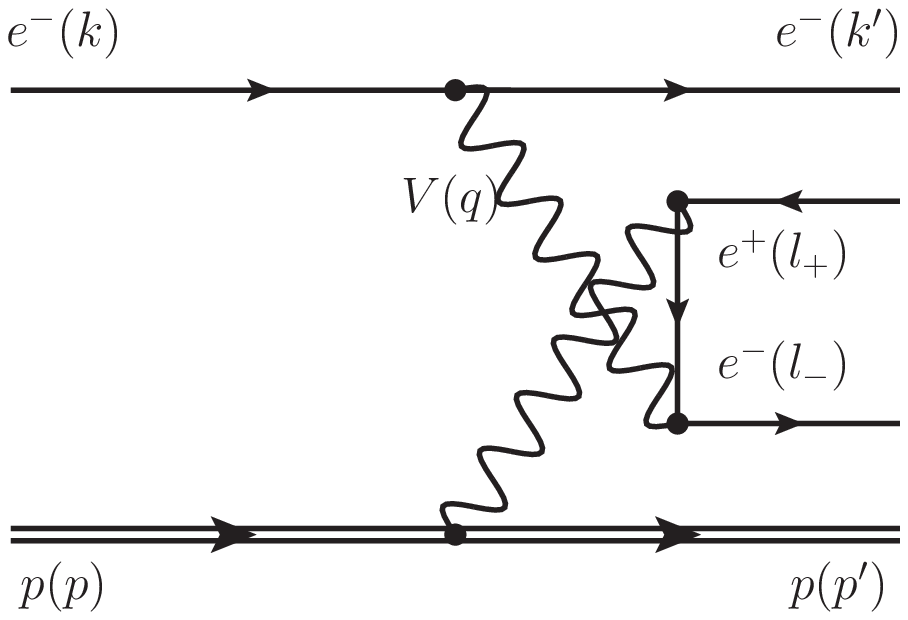}
\caption{Tree level Feynman diagrams contributing to the $e p \rightarrow e p l^+l^-$ amplitude.
Upper panel: exchange of the timelike boson $V$ and a spacelike $\gamma$ (TL).
Lower panel: the spacelike boson $V$ and a spacelike $\gamma$ (SL).
In addition to these direct (D) diagrams the exchange term (X), which consists of the same set of diagrams with scattered electron and electron of the $e^+e^-$ pair exchanged, also contributes.\label{fig:feyn_ep-epll}}
\end{figure}

The underlying diagrams for all fixed target experiments mentioned so far are shown in Fig.~\ref{fig:feyn_ep-epll}.
We calculate this process exactly in leading order of QED and furthermore apply leading order radiative corrections of the corresponding elastic scattering process to obtain an estimate of these corrections.\\
An electron beam of energy $E_0$ is scattered off a fixed target, which may either be a nucleon or a heavy nucleus of atomic numbers $(A,Z)$.
In the following, the target mass $M$ refers to the nucleon mass $M_N$ or to the mass of the heavy nucleus $M_A \simeq A \times M_N$.
As subprocess to the elastic scattering an intermediate vector particle $V$ is produced and creates a lepton pair ($l^+\,l^-$), where the lepton mass is denoted by $m_l$.
Although the existing and planned fixed target experiments only consider electron-positron pairs in which a bump hunt is performed, our calculations are performed generally for any kind of lepton species (i.e also applies to the $\mu^+\mu^-$ case), i.e. we do not neglect the mass of the lepton.\\
The isolated $\Ap$ production process is given by the coherent sum of diagrams (a) and (b) while the background, resulting from the exchange of a virtual photon, is given by the sum over all diagrams, where the intermediate vector particle $V$ in diagrams (a) and (b) is $\Ap$ and $\gamma^\ast$, respectively.\\
We assign a finite decay width $\Gamma_\Ap$ to the $\Ap$.
The partial decay width to a SM lepton pair $l^+l^-$ is given by
\[
  \Gamma_{\Ap\rightarrow l^+l^-} = \frac{ \alpha \varepsilon^2}{3 m_{\Ap}^2}\sqrt{m_\Ap^2-4m_l^2}\,(m_\Ap^2+2m_l^2),
\]
with $\alpha=e^2/(4\pi)\simeq 1/137$.\\
For kinematically forbidden decays to dark matter the total width can be related to the partial width by $\Gamma_\Ap=N_\text{eff}\times \Gamma_{\Ap\rightarrow l^+l^-}$, where $N_\text{eff}$ is a weight to account for other degrees of freedom in SM decays.
Since in this case the width is very small, only a small mass window around the peak will contribute to the signal and thus the cross sections for real and virtual $\Ap$ multiplied by $N_\text{eff}$ are equal \cite{Bjorken:2009mm,Freytsis:2009bh}.\\
In this work we denote the four-momenta of the initial and final beam electrons by $k=(E_0,\,\vec{k})$ and $k^\prime=(E_e^\prime,\,\vecp{k})$; the four-momenta of the initial and final target state by $p=(E_p,\,\vec{p})$ and $p^\prime=(E_p^\prime,\,\vecp{p})$ and the lepton pair four-momenta by $l_-=(E_-,\,\vec{l}_-)$ and $l_+=(E_+,\,\vec{l}_+)$, for the lepton and anti-lepton, respectively.
The initial and final electron spins are denoted by $s_k$ and $s_k^\prime$; the spins of the initial and final proton by $s_p$ and $s_p^\prime$; and the spins of the created lepton and anti-lepton by $s_-$ and $s_+$.
Furthermore we follow the conventions of Bjorken and Drell \cite{Bjorken:1964}.\\
The invariant amplitudes required to calculate the cross section can be read of from these Feynman diagrams.
As in the two diagrams in the upper panel of Fig.~\ref{fig:feyn_ep-epll} the intermediate boson $V$ is timelike, we refer to this amplitude as TL.
Correspondingly, we refer to the diagrams in the lower panel, where the $V$ is spacelike, as SL and their sum is denoted by SL + TL.\\
In the case that the $l^+l^-$ pair and the beam lepton are of the same species as for the existing experiments, another set of diagrams is allowed.
Since one cannot distinguish the electrons in the final state, the same diagrams of Fig.~\ref{fig:feyn_ep-epll} with the scattered (beam) electron and created electron of the pair exchanged, also have to be taken into account.
Therefore, following the notation of Ref.~\cite{Bjorken:1964}, we refer to the diagrams depicted in Fig.~\ref{fig:feyn_ep-epll} as ``direct'' contribution and to those with exchanged final state electrons as ``exchange'' contribution, labeled by $\D$ and $\X$, respectively.\\
For the TL diagrams one finds for the isolated $\Ap$ production process 
\begin{align}
 \mathcal{M}_{\Ap}^\text{TL} &\quad=\frac{i \,e^4\,\varepsilon^2}{\left(p^\prime-p\right)^2}\,\frac{-g^{\alpha \beta} + {q^{\prime \alpha}q^{\prime \beta} }/{m_{\Ap}^2}}{q^{\prime 2}-m_{\Ap}^2+i\,m_{\Ap}\,\Gamma_{\Ap}}\,J_N^\mu\,\mathcal{I}_{\mu\alpha}\,j^\text{pair}_\beta \label{epepll:eq_M_ab_Ap-1},
\end{align}
where $\Gamma_{\Ap}$ denotes the total $\Ap$ decay width.
The amplitude of the $\gamma^\ast$ background is given by:
\begin{align}
 \mathcal{M}_{\gamma^\ast}^\text{TL} &\quad=\frac{i \,e^4}{\left(p^\prime-p\right)^2}\,\frac{-g^{\alpha \beta}}{q^{\prime 2}}\,J_N^\mu\,\mathcal{I}_{\mu\alpha}\,j^\text{pair}_\beta, \label{epepll:eq_M_ab_gamma-1}
\end{align}
where the external momenta are denoted by $q=k-k^\prime$, $q^\prime=l_+ + l_-$ as on Fig.~\ref{fig:feyn_ep-epll}.
Furthermore the leptonic tensors are given by
\begin{align*}
\mathcal{I}_{\mu\alpha} &= \ubar_e(k^\prime,s_k^\prime) \left( \gamma_\mu \frac{ \fslash{(k-q^\prime)} +m}{\left(k-q^\prime \right)^2-m^2} \gamma_\alpha
+\gamma_\alpha \frac{ \fslash{(k^\prime+ q^\prime)} +m}{\left(k^\prime+q^\prime \right)^2-m^2} \gamma_\mu \right) u_e(k,s_k),\\
j^\text{pair}_\beta &= \ubar_l(l_-,s_-)\, \gamma_\beta\, v_l(l_+,s_+),
\end{align*}
with $m$ denoting the mass of the electron.
While in the case of a proton target the hadronic current $  J_N^\mu$ is given by
\[
   J_N^\mu =\ubar_N(p^\prime, s_p^\prime)\, \Gamma^\mu\, u_N(p,s_p),
\]
with the parametrization of $\Gamma_\mu (Q_t^2) \equiv F_1(Q_t^2)\, \gamma_\mu + F_2(Q_t^2)\,i\, \sigma_{\mu \nu} {q_t^\nu}/{2M}$ using the Dirac and Pauli form factors $F_1$ and $F_2$ and $Q_t={-(p-p^\prime)^2}>0$.
For a heavy nucleus it can be written to good approximation as
\[
 J_N^\mu = Z \cdot F(Q_t)\cdot (p+p^\prime)^\mu,
\]
where $F(Q_t)={3}/{\left(Q_t\,R\right)^3}\cdot \left({\sin{(Q_t\,R)}} - Q_t R\, \cos{(Q_t\,R)}\right)$ is the nuclear charge form factor with $R=1.21\,\mathrm{fm}\cdot A^{\frac{1}{3}}$. The nucleus spin as well as contributions from the breakup channel and nuclear excitations can be neglected to good approximation.
Effects due to the nucleus spin are suppressed by the large nucleus mass, which can be checked analytically.
The inelastic contribution can be neglected since the momenta transfered to the nucleus are small.\\
The numerator of the $\Ap$ propagator in Eq.~(\ref{epepll:eq_M_ab_Ap-1}) can be simplified as $(-g^{\alpha \beta})$ since the four-momentum $q^\prime$ is contracted with the lepton current $j^\text{pair}$ and thus the second term vanishes due to current conservation.\\
For the SL diagrams the invariant amplitude is given by
\begin{align}
 \mathcal{M}_{\gamma^\ast}^{SL} &\quad=\frac{i \,e^4}{\left(p^\prime-p\right)^2}\,\frac{-g^{\alpha \beta}}{q^2}\,J_N^\mu\,\tilde{\mathcal{I}}_{\mu\alpha}\,j^\text{beam}_\beta \label{epepll:eq_M_cd_gamma-1},
\end{align}
with
\begin{align*}
 \tilde{\mathcal{I}}_{\mu\alpha} &=  \ubar_l(l_-,s_-) \left( \gamma_\mu \frac{ \fslash{(q-l_+)} +m_l}{\left(q-l_+ \right)^2-m_l^2 } \gamma_\alpha + \gamma_\alpha \frac{ \fslash{(l_--q)} +m_l}{\left(l_- - q \right)^2-m_l^2} \gamma_\mu \right) v_l(l_+,s_+),\\
j^\text{beam}_\beta &= \ubar_e(k^\prime,s_k^\prime)\, \gamma_\beta\, u_e(k,s_k).
\end{align*}
Although the virtual $\Ap$ exchange via the SL process is not forbidden, it will not be considered here as it would not result in any bump in the $e^+e^-$ mass spectrum.
The propagator in Eq.~(\ref{epepll:eq_M_cd_gamma-1}) in that case would be replaced by
\begin{align*}
 \frac{-g^{\alpha \beta}}{q^2} \rightarrow \frac{-g^{\alpha \beta}}{q^2-m_{\Ap}^2},
\end{align*}
and due to the spacelike $q^2<0$ for scattering processes the denominator always leads to a suppression of this contribution, whereas the denominator in Eq.~(\ref{epepll:eq_M_ab_Ap-1}) leads to a peak in the signal.
Thus this contribution of virtual $\Ap$ exchange via the SL process to the cross section can be neglected.\\
\begin{figure}
\includegraphics[width=.48\linewidth]{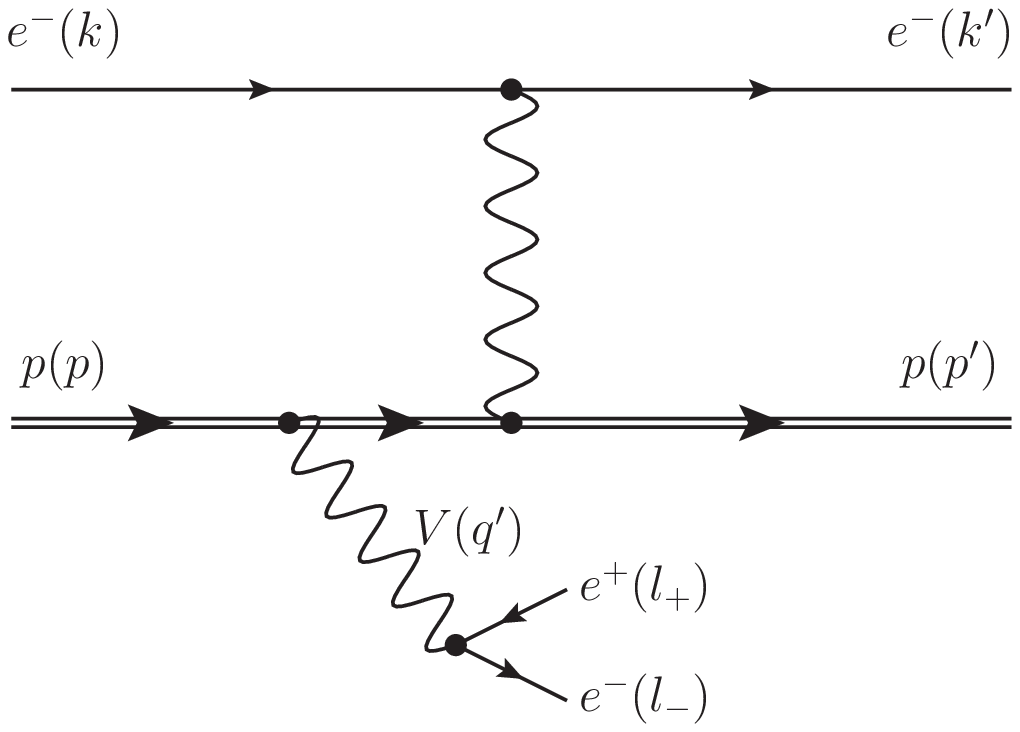} \includegraphics[width=.48\linewidth]{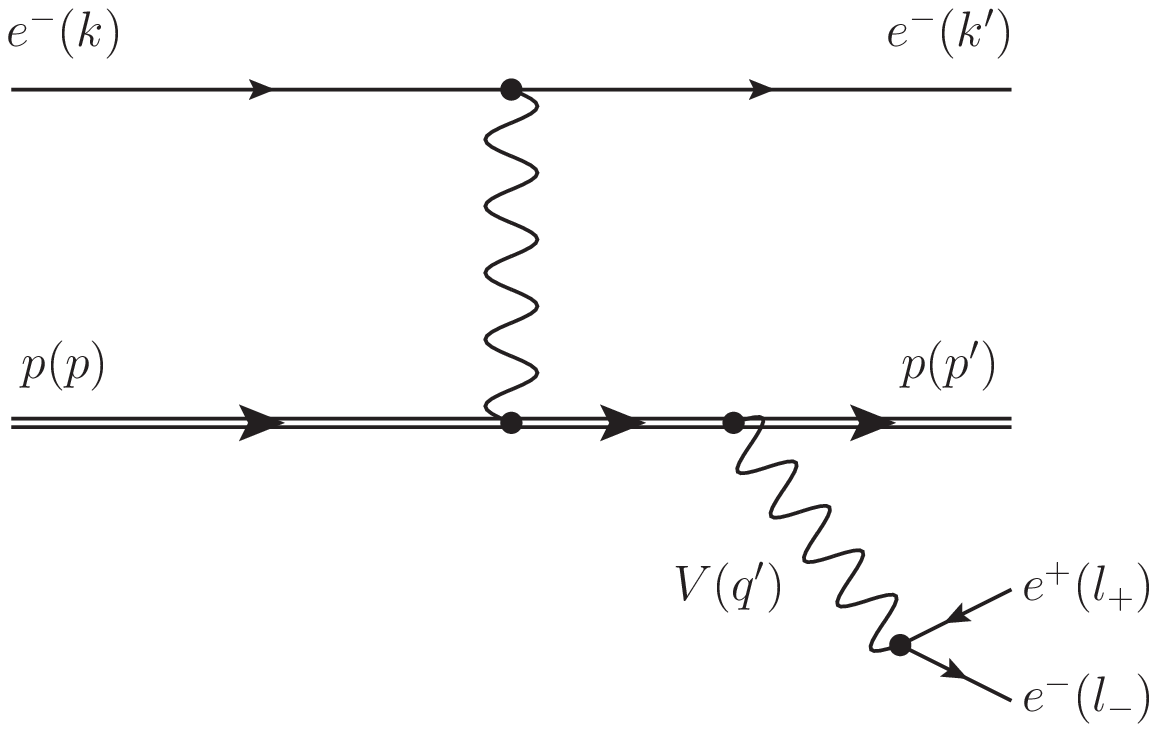}
\caption{Tree level Feynman diagrams of the double VCS contribution.\label{fig:feyn_ep-epll_vcs}}
\end{figure}
In the case of a proton target another important contribution, the double virtual Compton scattering (VCS), emerging from the third set of Feynman diagrams shown in Fig.~\ref{fig:feyn_ep-epll_vcs}, appears.
In case of a heavy nucleus target this term is strongly suppressed due to the large mass.
In this work we will restrict our study to estimate the influence of the nucleon pole contribution drawn in Fig.~\ref{fig:feyn_ep-epll_vcs} which serves as a good approximation.
The invariant amplitude is given by
\begin{align}
 \mathcal{M}_{\Ap}^\text{VCS} &\quad=\frac{-i \,e^4\,\varepsilon^2}{q^2}\,\frac{-g^{\alpha \beta} + {q^{\prime \alpha}q^{\prime \beta} }/{m_{\Ap}^2}}{q^{\prime 2}-m_{\Ap}^2+i\,m_{\Ap}\,\Gamma_{\Ap}}\,j_\text{beam}^\mu\,\mathcal{H}_{\mu\alpha}\,j^\text{pair}_\beta \label{epepll:eq_M_vcs_Ap-1},
\end{align}
for the isolated $\Ap$ production process, and 
\begin{align}
 \mathcal{M}_{\gamma^\ast}^\text{VCS} &\quad=\frac{-i \,e^4}{q^2}\,\frac{-g^{\alpha \beta}}{q^{\prime 2}}\,j_\text{beam}^\mu\,\mathcal{H}_{\mu\alpha}\,j^\text{pair}_\beta \label{epepll:eq_M_vcs_gamma-1}
\end{align}
for the $\gamma^\ast$ background, with
\begin{align*}
\mathcal{H}_{\mu\alpha} &= \ubar_p(p^\prime,s_p^\prime) \left( \Gamma_\mu(q_t + q^\prime) \frac{ \fslash{(p-q^\prime)} +M_N}{\left(p-q^\prime \right)^2-M_N^2} \Gamma_\alpha(-q^\prime) +\Gamma_\alpha(-q^\prime) \frac{ \fslash{(p^\prime+ q^\prime)} +M_N}{\left(p^\prime+q^\prime \right)^2-M_N^2} \Gamma_\mu(q_t + q^\prime) \right) u_p(p,s_p).
\end{align*}
As mentioned before, the electron from the scattered beam and the one from the lepton pair cannot be distinguished and besides the direct term the exchange term has to be accounted for.
Therefore the full amplitude of the process reads as
\begin{align}
 \mathcal{M}_{\Ap+\gamma^\ast} &= \left(\mathcal{M}_{\Ap}^\text{TL} + \mathcal{M}_{\gamma^\ast}^\text{TL} +\mathcal{M}_{\gamma^\ast}^\text{SL} \right) - \left(\left(\mathcal{M}_{\Ap}^\text{TL} + \mathcal{M}_{\gamma^\ast}^\text{TL} +\mathcal{M}_{\gamma^\ast}^\text{SL} \right)(e^-\leftrightarrow l^-)\right)\nonumber\\
 &= \left(\mathcal{M}_{\D,\,\Ap}^\text{TL} + \mathcal{M}_{\D,\,\gamma^\ast}^\text{TL} +\mathcal{M}_{\D,\,\gamma^\ast}^\text{SL} \right) - \left(\mathcal{M}_{\X,\,\Ap}^\text{TL} + \mathcal{M}_{\X,\,\gamma^\ast}^\text{TL} +\mathcal{M}_{\X,\,\gamma^\ast}^\text{SL} \right)
 \label{eq_M_A'+g_full},
\end{align}
for a heavy nucleus target and
\begin{align*}
 \mathcal{M}_{\Ap+\gamma^\ast} &= \left(\mathcal{M}_{\D,\,\Ap}^\text{TL} + \mathcal{M}_{\D,\,\gamma^\ast}^\text{TL} +\mathcal{M}_{\D,\,\gamma^\ast}^\text{SL}+\mathcal{M}_{\D,\,\gamma^\ast}^\text{VCS} \right) - \left(\mathcal{M}_{\X,\,\Ap}^\text{TL} + \mathcal{M}_{\X,\,\gamma^\ast}^\text{TL} +\mathcal{M}_{\X,\,\gamma^\ast}^\text{SL} +\mathcal{M}_{\X,\,\gamma^\ast}^\text{VCS} \right),
\end{align*}
for a proton target.
In the second term of Eq.~(\ref{eq_M_A'+g_full}) all quantities associated with the scattered electron and the pair electron are exchanged.
The exchange $\Ap$ term can be neglected, as the $\Ap$ propagator does not peak and thus a possible signal is suppressed by $\varepsilon^2$.
Due to the exchange of final state electron momenta, the amplitude describing the signal $\mathcal{M}_{\D,\,\Ap}^\text{TL}$ as well as the background contributions $\mathcal{M}_{\D,\,\gamma^\ast}^\text{TL}$ and $\mathcal{M}_{\X,\,\gamma^\ast}^\text{SL}$ contain a structure
\begin{align*}
 \frac{\fslash{(k-l_--l_+)}+m}{\left(k-l_--l_+\right)^2 - m^2},
\end{align*}
that contributes to the irreducible background.
This leads to a large contribution from $\mathcal{M}_{\X,\,\gamma^\ast}^\text{SL}$ in the case of forward scattering, since the denominator of the propagator is close to zero.
Forward scattering was proposed to enhance the signal strength, while not increasing the background $\mathcal{M}_{\gamma^\ast}^\text{SL}$.
Taking the background contribution $\mathcal{M}_{\X,\,\gamma^\ast}^\text{SL}$ into account, this argument is not applicable anymore, since now the background is also enhanced.\\
The cross section of the $e p \rightarrow e p e^+ e^-$ process is computed from the general expression for $2\rightarrow 4$ particle processes
\begin{align}
\begin{split}
 d \sigma&=\frac{1}{4\sqrt{(k \cdot p)^2-m^2M^2}}\,(2\pi)^4 
\delta^{(4)}\left(k+p-k^\prime-p^\prime-l_--l_+\right)\\
 &\quad\times \frac{d^3\vecp{k}}{(2\pi)^3\,2\, E_e^\prime}\, \frac{d^3\vecp{p}}{(2\pi)^3\,2\,E_p^\prime}\, \frac{d^3\vec{l_-}}{(2\pi)^3\,2\,E_-}\, \frac{d^3\vec{l_+}}{(2\pi)^3\,2\,E_+}\,\overline{\left|\mathcal{M}\right|^2}\label{eq_dcs-gen}.
\end{split}
\end{align}
Using a convenient set of variables we can express the cross sections as
\begin{align}
 \frac{d\sigma}{d m_{ll}\,d E_{e^\prime}^L\,d \Omega_{e^\prime}^L\,d\Absp{q}^\ast\,d\Omega_{q^\prime}^\ast\, d\Omega_+^{\ast \ast}} &= 
 \frac{\Absp{k}^L}{128\,\Abs{k}^L M}\frac{1}{\left(2\pi\right)^8}\frac{\lambda^\frac{1}{2}\!\!\left(s,\,M^2,\,m_{ll}^2 \right) \sqrt{m_{ll}^2-4m_l^2}}{2\,s}
 \,\overline{\left|\mathcal{M}\right|^2}\label{eq_dcs-recursive},
\end{align}
where $m_{ll}=\sqrt{q^{\prime 2}}$ is the invariant mass of the $l^+l^-$ pair, $s=\left(p+q \right)^2$ is the Mandelstam invariant of the $\gamma^\ast$-target subprocess, $\lambda\!\left(s,\,M^2,\,m_{ll}^2 \right) = \left(s-\left(M+m_{ll} \right)^2 \right)\left(s-\left(M-m_{ll} \right)^2 \right)$ denotes the kinematical triangle function.
Furthermore, we use the superscripts $()^L$ to label the lab frame, $()^\ast$ for the $(q+p)$ rest frame, and $()^{\ast \ast}$ for the $q^\prime$ rest frame.
This approach avoids ambiguities in the kinematics and automatically gives the full kinematically allowed region of the phase space.
In the considered type of experiments only a small fraction of the kinematically allowed phase space is probed.
The allowed region is given by the detector acceptances in the lab frame.
Therefore it is convenient to calculate the cross section directly in terms of lab frame quantities and to use the recursively built up phase space as a cross-check.\\
Since fixed target experiments are considered here, the target four-momentum $p$ simplifies to $p=(M,\,\vec{0})$.
Furthermore in the considered experiments the detectors and the beam are aligned in the same plane which we account for by the choice of our parametrization of the momentum vectors of the detected particles.
Since neither the scattered hadron nor the scattered electron will be detected in the experiments, as long as the electrons are treated as distinguishable particles, the dependence of the cross section on their four-momenta has to be eliminated.
Therefore the three-momentum conserving $\delta$-function is used to eliminate the three-momentum of the final hadron state $\vecp{p}$ and energy conservation is used to express the absolute value of the three-momentum of the scattered electron $\Absp{k}$.
The remaining dependence of the cross section on the electron scattering angle is removed by integration over the full solid angle $\Omega_{e^\prime}$.
Furthermore one is interested in the cross section as function of the invariant mass of the created lepton pair, which is equal to the squared four-momentum of the intermediate vector boson $q^{\prime 2}=m_{ll}^2$.
Therefore we trade the absolute value of $\vec{l}_-$ for $q^{\prime 2}$.\\
Thus one finds from Eq.~(\ref{eq_dcs-gen}) for the differential cross section in the lab frame
\begin{align}
\frac{d \sigma}{d\Abs{l_+} \,d\Omega_+\,d\Omega_-\,d\Omega_{e^\prime}\,dq^{\prime\,2}}
&=\frac{1}{128\,\Abs{k}\,M} \frac{1}{\left(2\pi\right)^8}\frac{\Absp{k}^2\Abs{l_+}^2\Abs{l_-}^2} {E_{p^\prime}E_{k^\prime}E_{A^\prime}E_{+}E_{-}}
\left(\left| \frac{\partial \delta_1}{\partial \Absp{k}}\right|\left| \frac{\partial \delta_2}{\partial \Abs{l_-}} \right| \right)^{-1}\overline{\left|\mathcal{M}\right|^2},
\label{eq_epepll:dxs8-1}
\end{align}
where this equation is understood to be evaluated with $\Abs{l_+}$ and $\Absp{k}$ given in Eqs.~(\ref{eq:epepll_norm_l-final}) and (\ref{eq:epepll_norm_kp_final}), and $\frac{\partial \delta_1}{\partial \Absp{k}}$ and $\frac{\partial \delta_2}{\partial \Abs{l_-}}$ are given by Eqs.~(\ref{eq_epepll:ddelta1}) and (\ref{eq_epepll:ddelta2}), respectively.
A more detailed derivation of the cross section is presented in Appendix~\ref{app_xs-derivation}.\\
Furthermore, we will apply radiative corrections of elastic electron-proton scattering to the cross section to achieve a better comparability with the experimental data.
Therefore the cross section of Eq.~(\ref{eq_epepll:dxs8-1}) is multiplied by Eq.~(A71) of Ref.~\cite{Vanderhaeghen:2000ws}.
By applying these radiative corrections the value of the cross section is reduced by an amount in the range of $10 - 20\, \%$.\\
The comparison with experimental data can be performed by integrating Eq.~(\ref{eq_epepll:dxs8-1}) over the experimental acceptances.
To obtain the acceptance integrated cross section $\Delta\sigma$, which can be related to experimental count rates by multiplication with the luminosity, a non-trivial 8-fold integration is necessary.
Furthermore the structure of the squared matrix element contains several strongly peaked structures which makes the numerical calculation of this integral challenging.
Any of the fermion and photon propagators in the Feynman diagrams shown in Fig.~\ref{fig:feyn_ep-epll} can possibly be near the mass shell in a certain kinematical setting.
Although there is no real divergence existing, since the non-vanishing mass of the electron serves as a regulator, the calculation of this strongly peaked structures either needs further approximations or a large numerical effort.
In our study we try to use as less approximations as possible.
We thus decide to use an integration method that allows to deal with these peaked structures by increasing the numerical precision.
Therefore for the numerical integration the VEGAS algorithm \cite{Lepage:1977sw} has been chosen, which is a well established Monte Carlo integration method in particle physics.
The standard deviation and the $\chi^2$ of the result of the integration are used to decide whether the computed value is reasonable or not.
During our calculations it turned out, that - at least for the case of MAMI kinematics -  one cannot use a vanishing electron mass to achieve numerically stable results.\\
In order to perform these calculations in a reasonable amount of time, we have performed a highly parallelized calculation.
Therefore the integral is computed on Graphics Processing Units (GPUs) using the NVIDIA CUDA framework \cite{nvidia} and the implementation of the VEGAS algorithms on GPUs published in Ref.~\cite{Kanzaki:2010ym}.
The use of the GPU version reduces the time needed for the evaluation of the acceptance integrated cross section by a factor of $\sim 60$.
We have checked the results achieved by the GPU calculation with ordinary calculations on CPUs and find that for a same numerical precision the results are equal within their standard deviations, which are below $10^{-4}$ relative to the obtained value.\\
The radiative background is described by the acceptance integrated cross section
\begin{equation}
 \Delta\sigma_\gamma \propto \left| \left(\mathcal{M}_{\gamma^\ast}^\text{TL} + \mathcal{M}_{\gamma^\ast}^\text{SL}\right) - \left(\left(\mathcal{M}_{\gamma^\ast}^\text{TL} + \mathcal{M}_{\gamma^\ast}^\text{SL}\right)(e^- \leftrightarrow l^-)\right) \right|^2 \label{eq_epepll:xs-radbkg-1},
\end{equation}
where the prefactors on the right-hand side are the same as appearing in Eq.~(\ref{eq_epepll:dxs8-1}).
For later use, besides the cross section of the process including $\gamma^\ast$ and $\Ap$, we define the direct timelike radiative background cross section and the direct timelike $\Ap$ cross section as
\begin{align}
  \Delta\sigma_{\Ap+\gamma} &\propto \left| \mathcal{M}_{\D+\X,\,\Ap}^\text{TL} + \mathcal{M}_{\D+\X,\,\gamma^\ast}^\text{TL} +  \mathcal{M}_{\D+\X,\,\Delta\gamma^\ast}^\text{SL} \right|^2 \label{eq_epepll:xs-full-1},\\
 \Delta\sigma_\gamma^\text{TL} &\propto \left| \mathcal{M}_{\D,\,\gamma^\ast}^\text{TL}\right|^2 \label{eq_epepll:xs-radbkg-tl-1},\\
 \Delta\sigma_{\Ap} &\propto \left| \mathcal{M}_{\D,\,\Ap}^\text{TL}\right|^2 \label{eq_epepll:xs-A'-tl-1},
\end{align}
respectively.\\
In order to compute exclusion limits on the coupling strength $\varepsilon$ from existing data, a relation between the cross sections of Eqs.~(\ref{eq_epepll:xs-radbkg-1}) and (\ref{eq_epepll:xs-A'-tl-1}) giving rise to $\varepsilon$ is required.
We split the $\Ap + \gamma$ cross section as
\[
 \Delta\sigma_{\Ap+\gamma} = \Delta\sigma_\gamma + \Delta\sigma_{\Ap} + \Delta\sigma_\text{int},
\]
with $\Delta\sigma_\text{int}$ denoting the interference part.
Dividing Eq.~(\ref{eq_epepll:xs-full-1}) by Eq.~(\ref{eq_epepll:xs-radbkg-1}) leads to
\[
 \frac{\Delta\sigma_{\Ap + \gamma}}{\Delta\sigma_{\gamma}} = 1 + \frac{3\pi}{2N} \frac{\varepsilon^2}{\alpha} \frac{m_{\Ap}}{\delta m} \,\frac{\Delta\sigma_\gamma^\text{TL}}{\Delta\sigma_{\gamma}} + \frac{\Delta\sigma_\text{int}}{\Delta\sigma_{\gamma}}.
\]
We have used Eq.~(19) of Ref.~\cite{Bjorken:2009mm} in order to approximate the ratio of $\sigma_{\Ap}$ and $\sigma_\gamma^\text{TL}$ as
\[
\frac{\Delta\sigma_{\Ap}}{\Delta\sigma_\gamma^\text{TL}} =  \frac{3\pi}{2N} \frac{\varepsilon^2}{\alpha} \frac{m_{\Ap}}{\delta m},
\]
where $N$ is the ratio of the decay widths $\Gamma_{\Ap\rightarrow e^+e^-}$ and $\Gamma_{\Ap\rightarrow \mu^+\mu^-}$ taking other possible final states into account and $\delta m$ is the experimental mass resolution, i.e. the mass bin width.
For $\Ap$ masses $\gtrsim 400\,\MeV$ hadrons also contribute to the final state and thus our parametrization of $N$ is not valid anymore.
In the $\Ap$ mass range considered in this work only electrons and muons are contributing as possible final states.
Our numerical calculations for a wide range of parameters $m_{\Ap}$ and $\varepsilon$ of the interference part $\sigma_\text{int}$ from the cross sections (\ref{eq_epepll:xs-radbkg-1}), (\ref{eq_epepll:xs-full-1}), and (\ref{eq_epepll:xs-A'-tl-1}) show that the interference between $\Ap$ signal and QED background can be neglected.
We find, that ${\Delta\sigma_\text{int}}/ {\Delta\sigma_{\gamma}}$ is less than $10^{-3}$, which is in the range of the achieved numerical precision.
Furthermore we find a very good agreement of the approximated ${\sigma_{\Ap}}/{\sigma_\gamma^\text{TL}}$ with our exact calculation for the largest part of the parameter region for $m_{\Ap}$ and $\varepsilon$.
Therefore $\varepsilon$ can be computed from the cross section ratio as
 \begin{equation}
  \varepsilon^2 = \left(\frac{\Delta\sigma_{\Ap + \gamma}}{\Delta\sigma_{\gamma}} -1 \right)\, \frac{\Delta\sigma_\gamma}{\Delta\sigma_\gamma^\text{TL}}\, \frac{2\,N\,\alpha}{3\pi}\, \frac{\delta m}{m_{\Ap}}.
\label{epsextr_eq:epssq}
 \end{equation}
The ratio ${\Delta\sigma_{\Ap + \gamma}}/{\Delta\sigma_{\gamma}}$ is the (aimed) signal sensitivity, which has to be determined from the experiment.
Furthermore, by using the ratio ${\Delta\sigma_{\Ap + \gamma}}/{\Delta\sigma_{\gamma}}$ for the extraction of $\varepsilon^2$, possible effects not accounted for in our approximation of the nuclear current will cancel each other.
For the prediction of exclusion limits we estimate ${\Delta\sigma_{\Ap + \gamma}}/{\Delta\sigma_{\gamma}} - 1$ as signal over background ratio
\[
 \frac{\sqrt{\#S}}{\#B} = \frac{2}{\sqrt{\Delta\sigma_{\gamma}\times L}},
\]
where $\#S$ and $\#B$ are the numbers of signal and background events in one mass bin, respectively, and L is the integrated luminosity.
The factor $2$ results from the fact, that in agreement with other publications we determine the exclusion limits on the $2\,\sigma$ level.
Since the exclusion limit on the coupling strength $\varepsilon^2$ is depending linearly on the ratio of the background cross section $\Delta\sigma_\gamma$ to the TL cross section with distinguishable final state electrons $\Delta\sigma_\gamma^\text{TL}$, the precise knowledge of these quantities is crucial to obtain an accurate result.
Therefore the next section will deal with the analysis of these background ratios for the existing experiments.
 
\section{Comparison of experimental data and theory calculations for MAMI\label{sec:MAMI_results}}
Two dedicated fixed target experiments, one by the A1 collaboration at MAMI \cite{Merkel:2011ze} and the APEX experiment at JLAB \cite{Abrahamyan:2011gv}, have already started taking data.
\subsection{Test run 2010}
\begin{table}
 \begin{tabular}{|l|c|c|c|}
  \hline
 & momentum & horizontal angle & vertical angle\\
 \hline
 A & $\pm 10  \%$ & $\pm 75$ mrad & $\pm 70$ mrad\\
  \hline
 B & $\pm 7.5 \%$ & $\pm 20$ mrad & $\pm 70$ mrad\\
  \hline
 \end{tabular}
\caption{Acceptances of the used spectrometers A and B at MAMI \cite{Blomqvist:1998xn}.\label{tab:MAMI_acceptance}}
\end{table}

\begin{figure}
 \includegraphics[angle=-90,width=.5\linewidth]{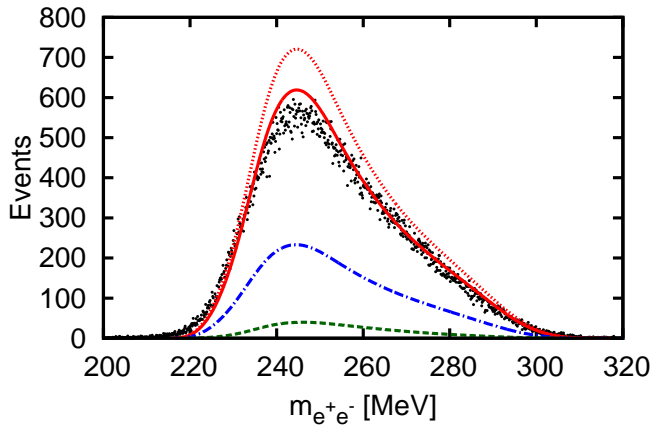}
 \caption{Comparison of theory calculations and experimental data for a $m_{e^+e^-}$ bin width of $0.125\,\MeV$.
Black points: Data taken in a particular run of the MAMI 2010 experiment \cite{Merkel:2011ze} in setup 1.
Solid curve: Theory calculation of the background cross section.
Dotted curve: Theory calculation of the background cross section without radiative corrections.
Dashed-dotted curve: Theory calculation of the direct SL + TL cross section.
Dashed curve: Theory calculation of the direct TL cross section.}
 \label{fig:mami2010_comparison_expvsth}
\end{figure}

\begin{figure}
\includegraphics[width=.4\linewidth,angle=-90]{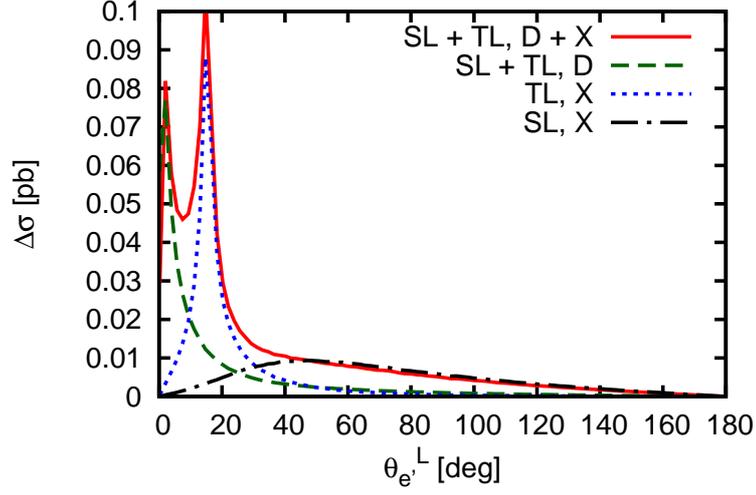}
\caption{Angular distribution per $0.5^\circ$ with respect to the polar angle of the scattered electron for the MAMI 2010 experiment.
\label{fig:mami2010_angular}} 
\end{figure}

\begin{figure*}
\begin{tabular}{cc}
  \includegraphics[angle=-90,width=.48\linewidth]{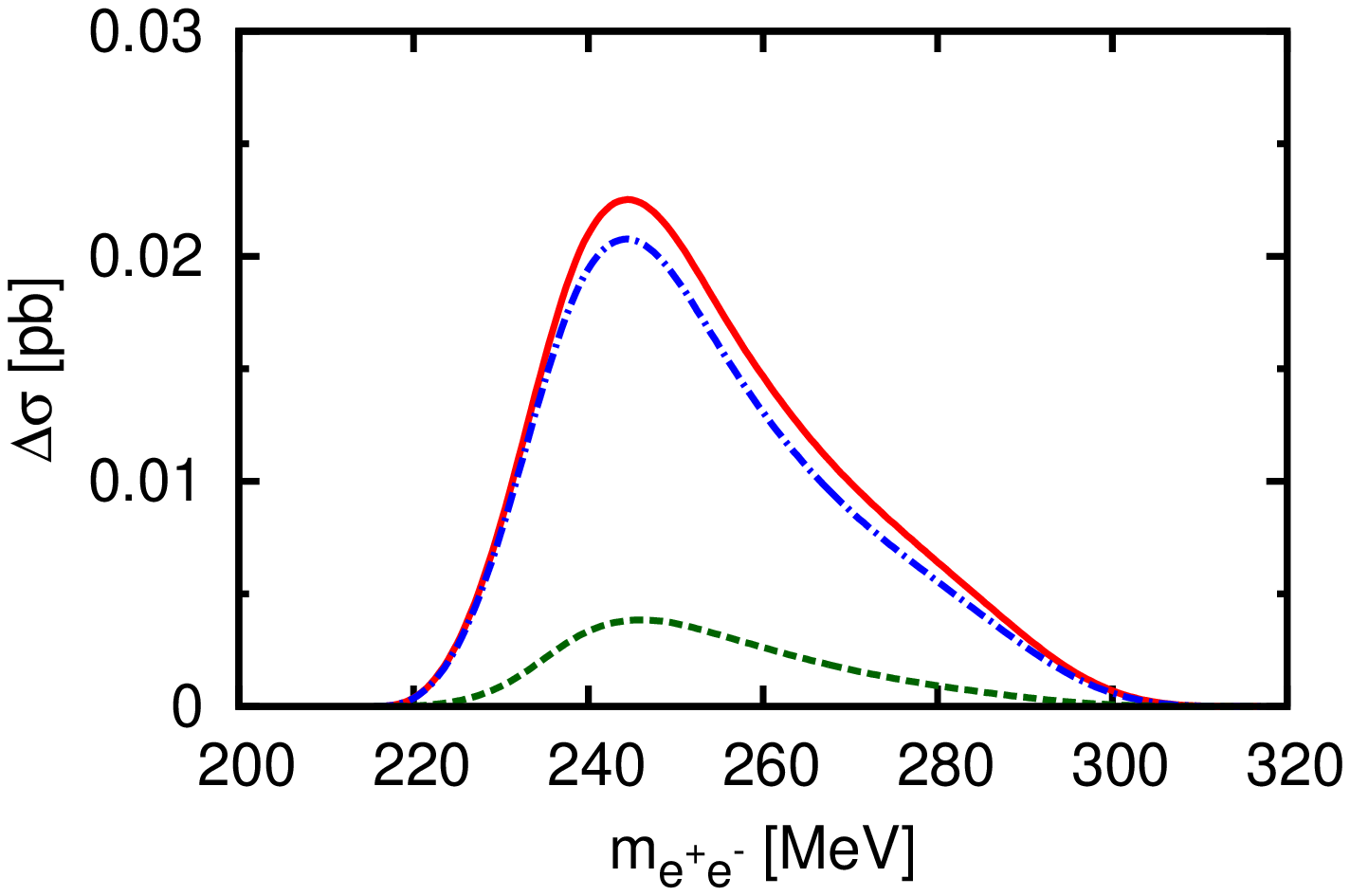} &
  \includegraphics[angle=-90,width=.48\linewidth]{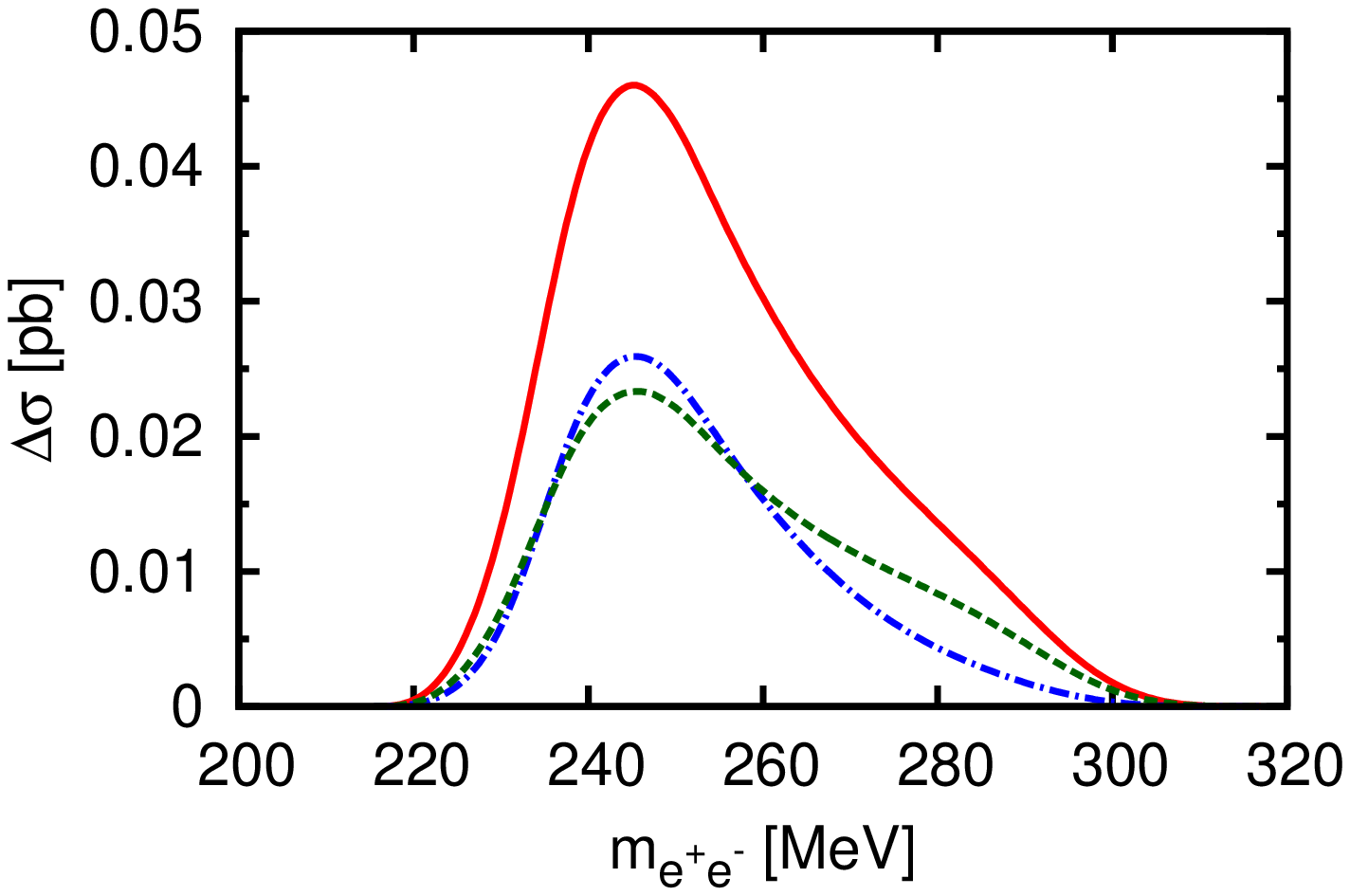}
\end{tabular}
 \caption{Calculated direct (left panel) and exchange (right panel) term of the cross section assuming distinguishable electrons in the final state.
Solid curve: SL + TL cross section.
Dashed curve: TL.
Dashed-dotted curve: SL}
 \label{fig:mami2010_calc_noASvsAS}
\end{figure*}

\begin{figure}
 \includegraphics[angle=-90,width=.5\linewidth]{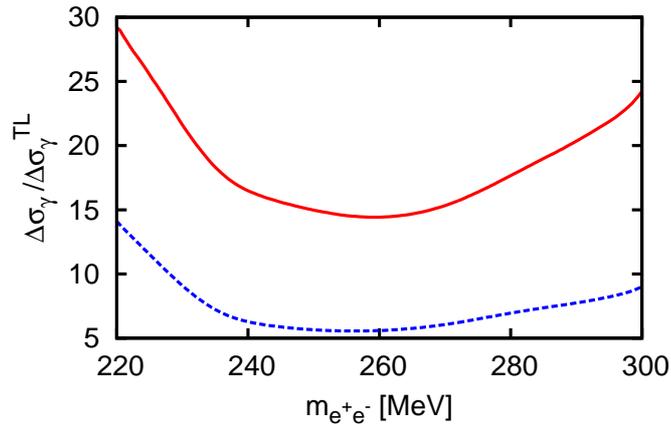}
 \caption{Solid (dashed) curve: Ratio of the background cross section $\Delta\sigma_{\gamma,\,\D+\X}$ ($\Delta\sigma_{\gamma,\,\D}$) to the direct TL cross section $\Delta\sigma_\gamma^\text{TL}$.}
 \label{fig:mami2010_ratio_fullvsab}
\end{figure}

A first test run to proof the feasibility of a dedicated $\Ap$ fixed target search experiment has been performed at MAMI by the A1 Collaboration in 2010 \cite{Merkel:2011ze}.
In this experiment no evidence for the existence of the $\Ap$ could be found and an exclusion limit on the $\Ap$ parameter space was formulated.
A sample of the data taken in this experiment compared to our calculations can be seen in Fig.\ref{fig:mami2010_comparison_expvsth}.\\
The kinematical settings of this experiment can be taken from Table 1 in Ref.~\cite{Merkel:2011ze}.
For the comparison of the calculation and the data, the setup 1 as given in Ref.~\cite{Merkel:2011ze} was chosen, since for this setup a luminosity measurement has been performed, finding an integrated luminosity of $\mathcal{L} = 41.4\, \text{fb}^{-1}$ for the selected sample of events.
A background contribution of around $5\%$ was already subtracted in this sample, the systematic uncertainty in the luminosity from the knowledge of the thickness of the target foil is below $5\%$.
The acceptances as shown in Table~\ref{tab:MAMI_acceptance} have been used as integration limits for the theory calculation.
Unless mentioned otherwise, the $m_{ll}$ integration is performed over a range of $0.5\,\MeV$, which is equal to the typical FWHM mass resolution of the considered experiments.\\
As seen on Fig.~\ref{fig:mami2010_comparison_expvsth}, our calculation (solid curve) of the radiative background given by Eq.~(\ref{eq_epepll:xs-radbkg-1}) and the experimental data (points) are in good agreement.
Due to our estimate of the nuclear current and of the radiative corrections we expect the small discrepancy between theory and data seen from Fig.~\ref{fig:mami2010_comparison_expvsth}.
The influence of the radiative corrections is displayed by the solid and dotted curve on Fig.~\ref{fig:mami2010_comparison_expvsth} which are calculated with and without radiative corrections, respectively.
It is obvious from Fig.~~\ref{fig:mami2010_comparison_expvsth}, that the applied radiative corrections lower the result of the theory calculation by an amount in the range of $10 - 20\, \%$, as mentioned in section~\ref{sec:calculations}.
The calculation of the full QED radiative corrections for such a process is very involved.
However, one can see from Fig.~\ref{fig:mami2010_comparison_expvsth}, that our approximate treatment of the radiative corrections already provides a very good approximation, as theory and data already are in good agreement.\\
The dashed (dashed-dotted) curve shows the direct TL (SL + TL) cross section.
This indicates, that a large contribution to the cross section results from the antisymmetrization due to the indistinguishability of the scattered beam electron and the pair electron.
The kinematical setting has been optimized to reduce the SL background.\\
The angular distribution with respect to the polar angle of the scattered electron presented in Fig.~\ref{fig:mami2010_angular} points out, that for the 2010 A1 experiment the crossed TL amplitude is responsible for a second peak in the background cross section compared to the direct amplitude (dashed curve) which only peaks at very forward scattering followed by a rapidly dropping tail.
The exchange SL term nevertheless enhances the tail of the angular distribution significantly.\\
Fig.~\ref{fig:mami2010_calc_noASvsAS} reveals, that in the chosen kinematic setting the exchange term contribution is about twice as large as the direct SL part, which initially should be minimized.
This means, that the largest contribution to the radiative background does not originate as assumed from the processes given by the direct SL Feynman diagrams of Fig.~\ref{fig:feyn_ep-epll}, but from the processes described by diagrams with exchanged final state electrons.\\
For the investigated kinematic setting we calculate the ratio of the background cross section to the direct TL cross section which is the crucial quantity entering the determination of the exclusion limit on $\varepsilon^2$, according to Eq.~(\ref{epsextr_eq:epssq}).
One notices from Fig.~\ref{fig:mami2010_ratio_fullvsab} (solid curve) that the ratio $\Delta\sigma_{\gamma,\,\D+\X}/\Delta\sigma_\gamma^\text{TL}$ smoothly varies between $15$ and $25$ for most of the invariant mass range.
Neglecting the necessary contribution of the exchange term to the cross section, the ratio is lower by a factor of about $3$ for the investigated range (dashed curve on Fig.~\ref{fig:mami2010_ratio_fullvsab}).
\subsection{2012\label{sec:MAMI2012}}
\begin{table}
 \begin{tabular}{|c|c|c|c|}
  \hline
  & $E_0\, [\text{MeV}]$ & $\Abs{l}_+\, [\text{MeV}]$ & $\Abs{l}_-\, [\text{MeV}]$\\
  \hline
  kin057 & 180 & 78.7  & 98\\
  kin072 & 240 & 103.6 & 132.0\\
  kin077 & 255 & 110.1 & 140.4\\
  kin091 & 300 & 129.5 & 164.5\\
  kin109 & 360 & 155.4 & 197.6\\
  kin138 & 435 & 190.7 & 247.7\\
  kin150 & 495 & 213.7 & 271.6\\
  kin177 & 585 & 250.0 & 317.3\\
  kin218 & 720 & 309.2 & 392.7\\
  \hline
 \end{tabular} 
\caption{Kinematics of the MAMI 2012 $\Ap$ search\label{tab:mami2012_kins}.
Electron scattering angle: $\phi_{-}=20.01^\circ$ (spectrometer A).
 Positron scattering angle: $\phi_{+}=-15.63^\circ$ (spectrometer B).
The number in the label of the kinematics refers to the invariant mass around which a setting is centered.}
\end{table} 

\begin{figure*}
 \includegraphics[angle=-90,width=.32\linewidth]{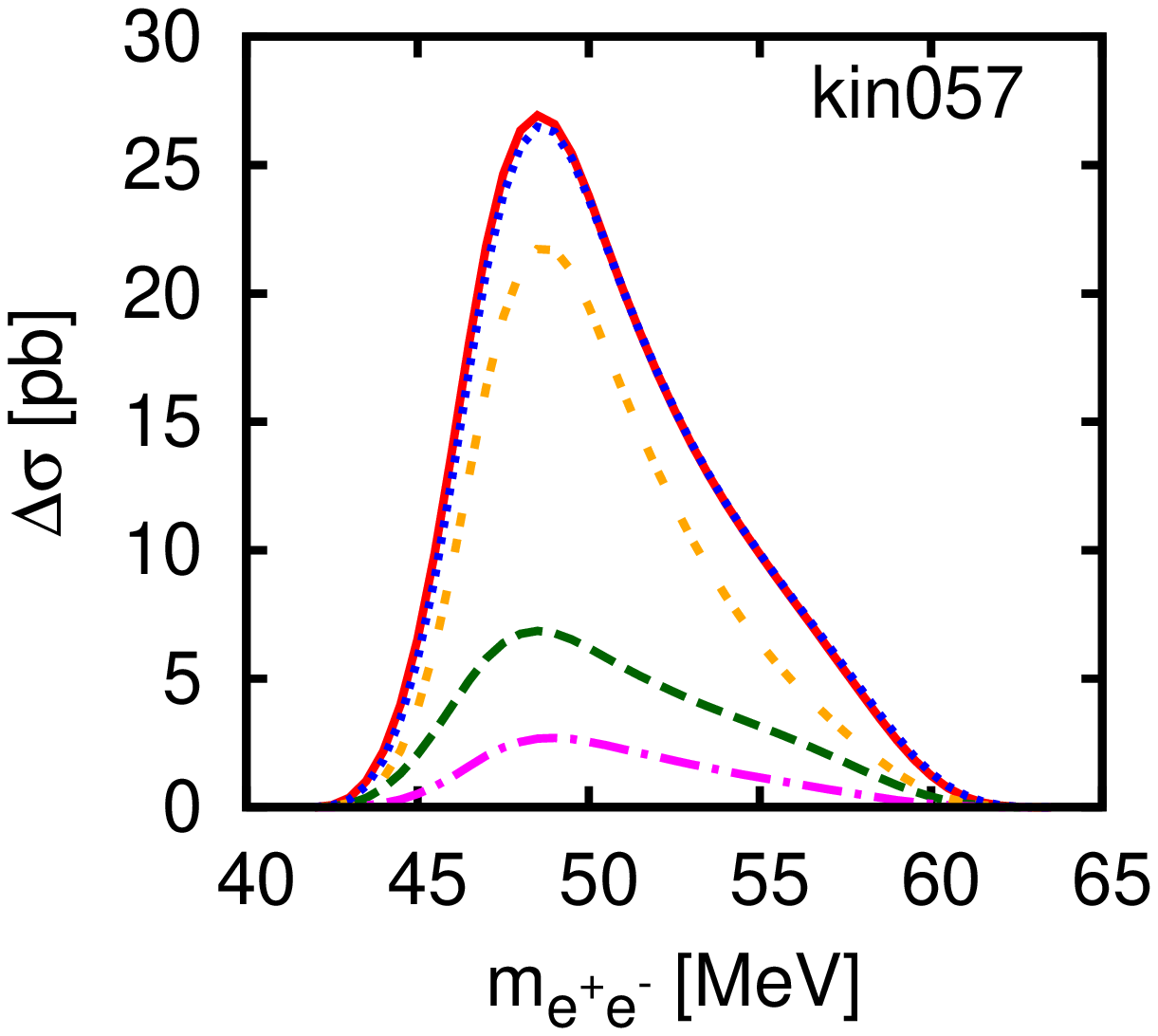}
 \includegraphics[angle=-90,width=.32\linewidth]{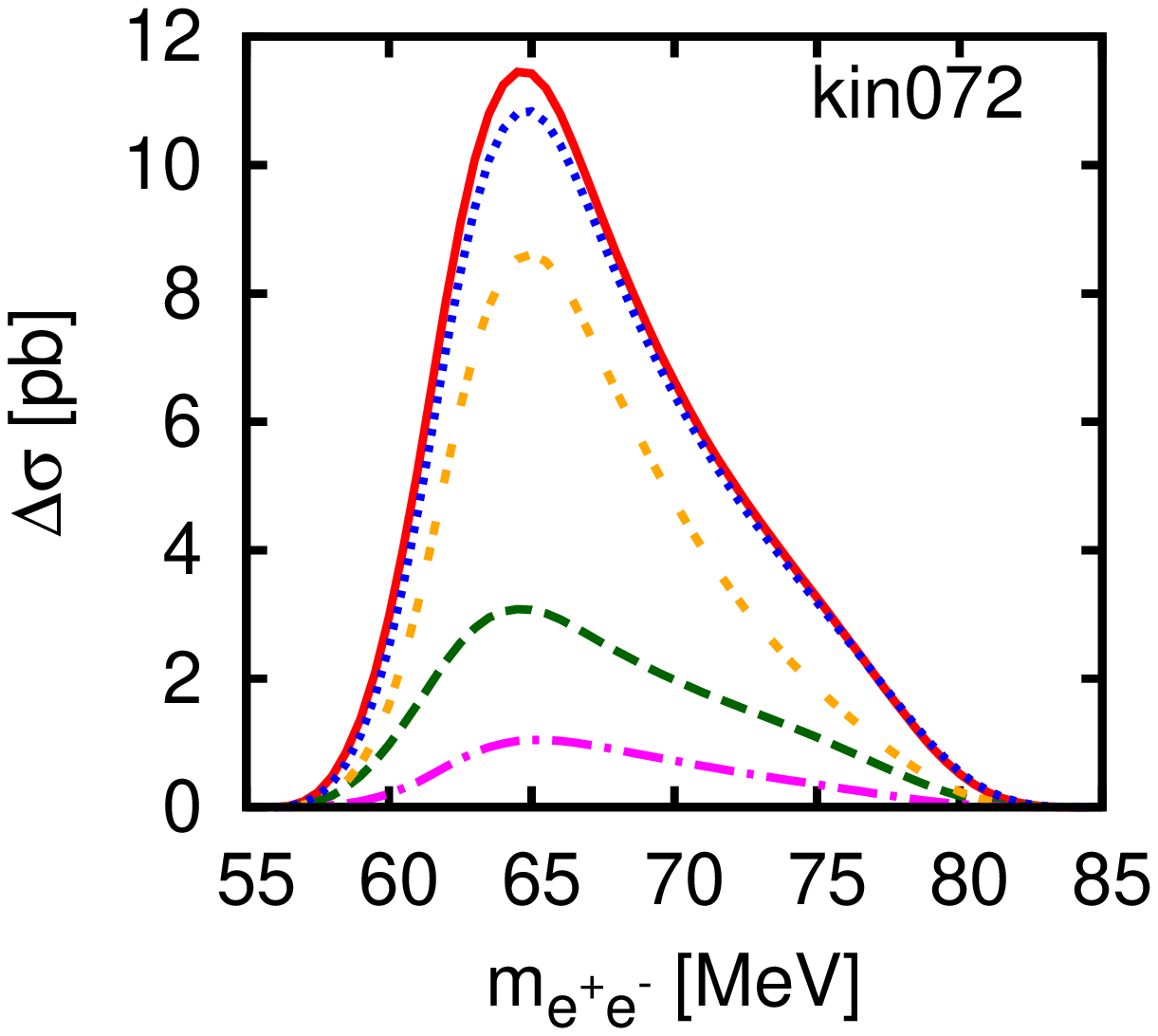}
 \includegraphics[angle=-90,width=.32\linewidth]{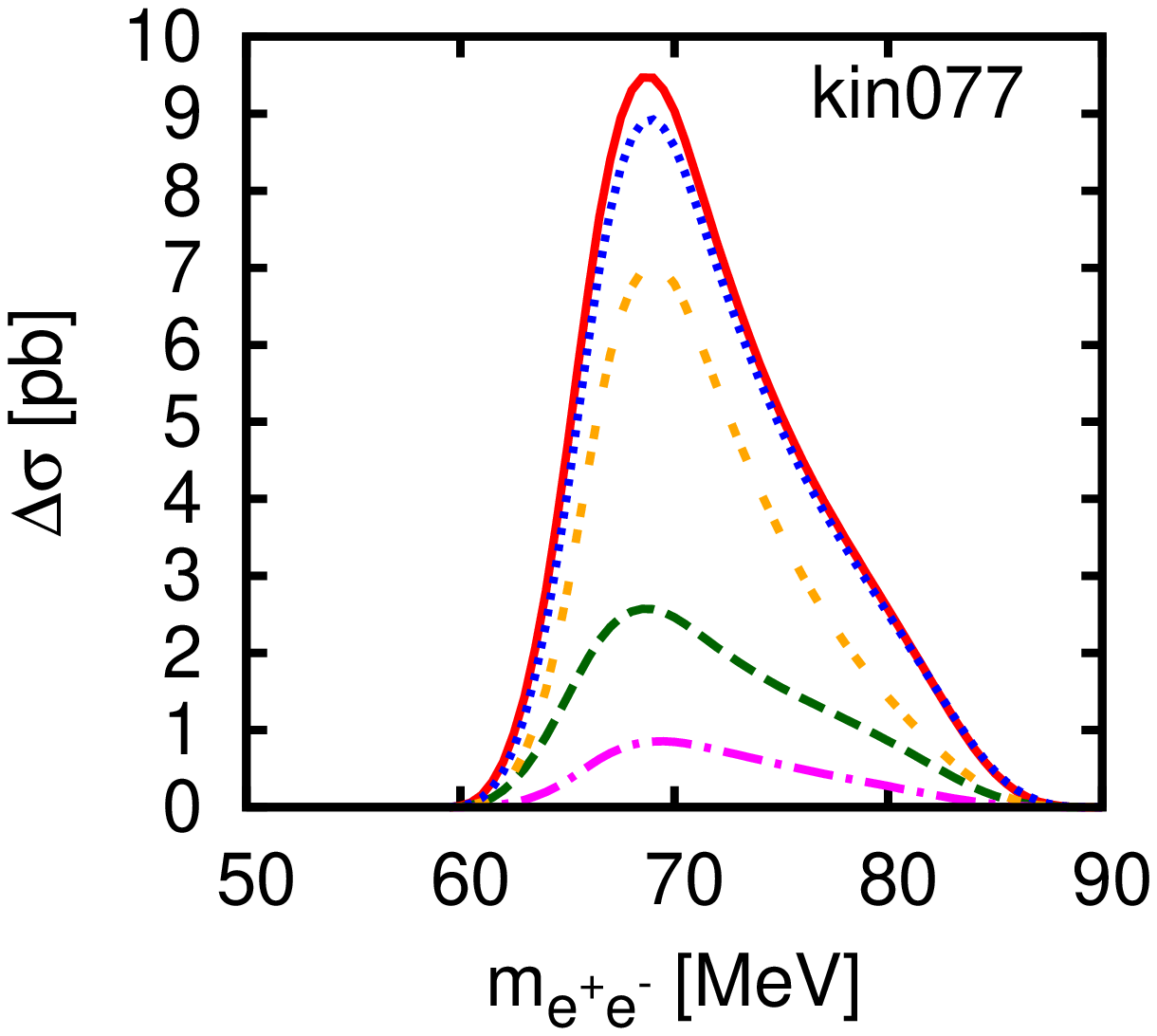}\\
 \includegraphics[angle=-90,width=.32\linewidth]{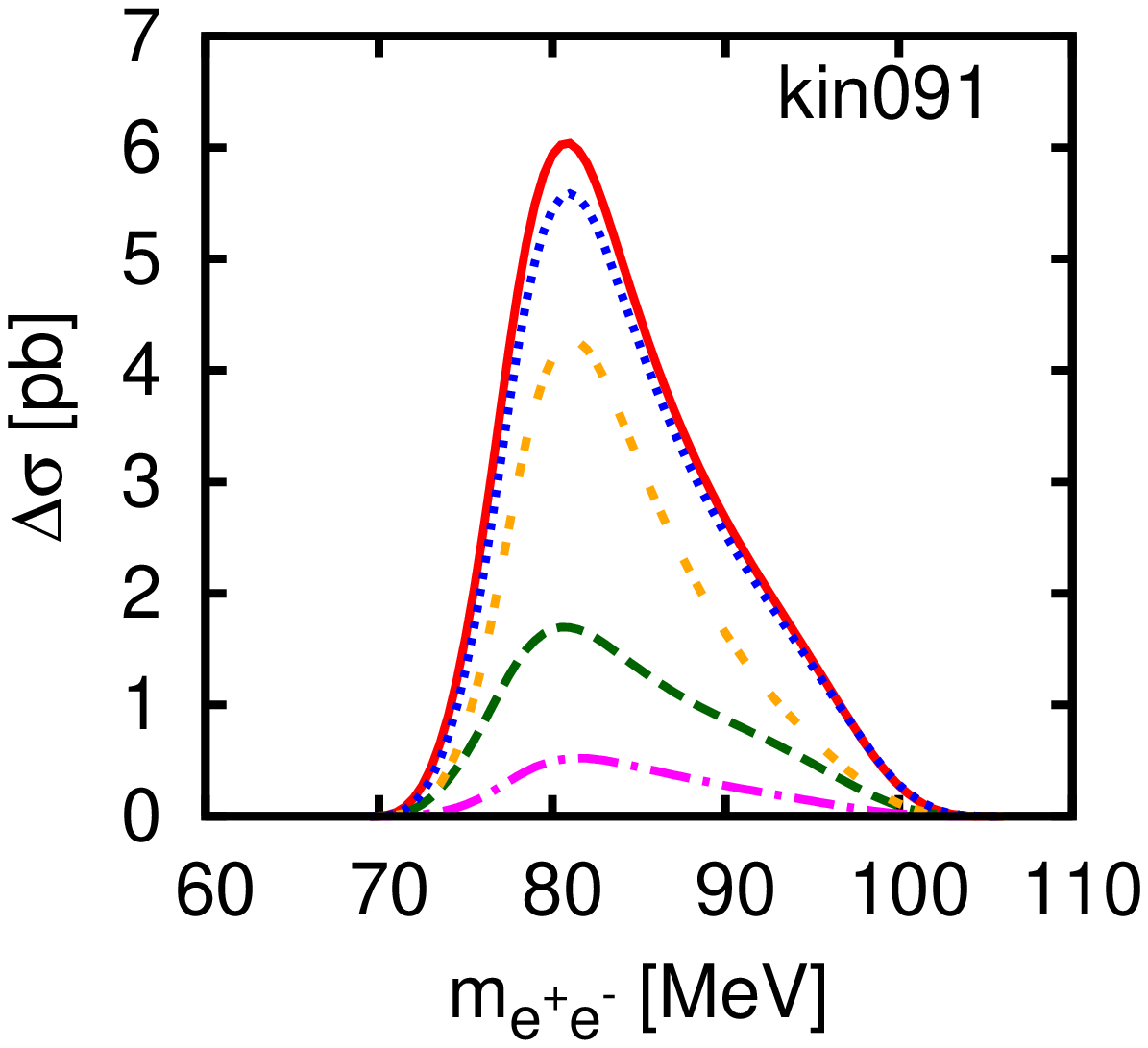}
 \includegraphics[angle=-90,width=.32\linewidth]{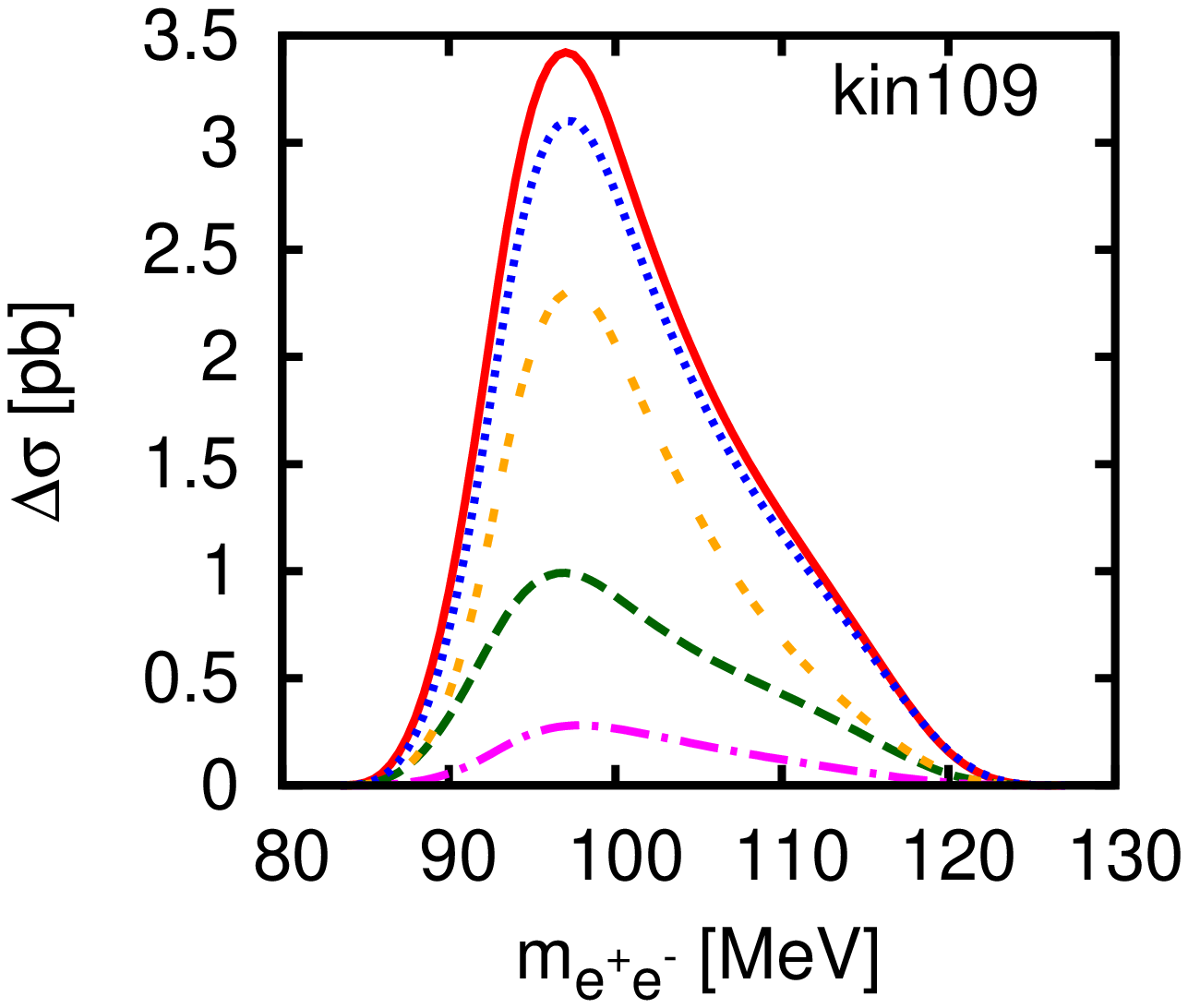}
 \includegraphics[angle=-90,width=.32\linewidth]{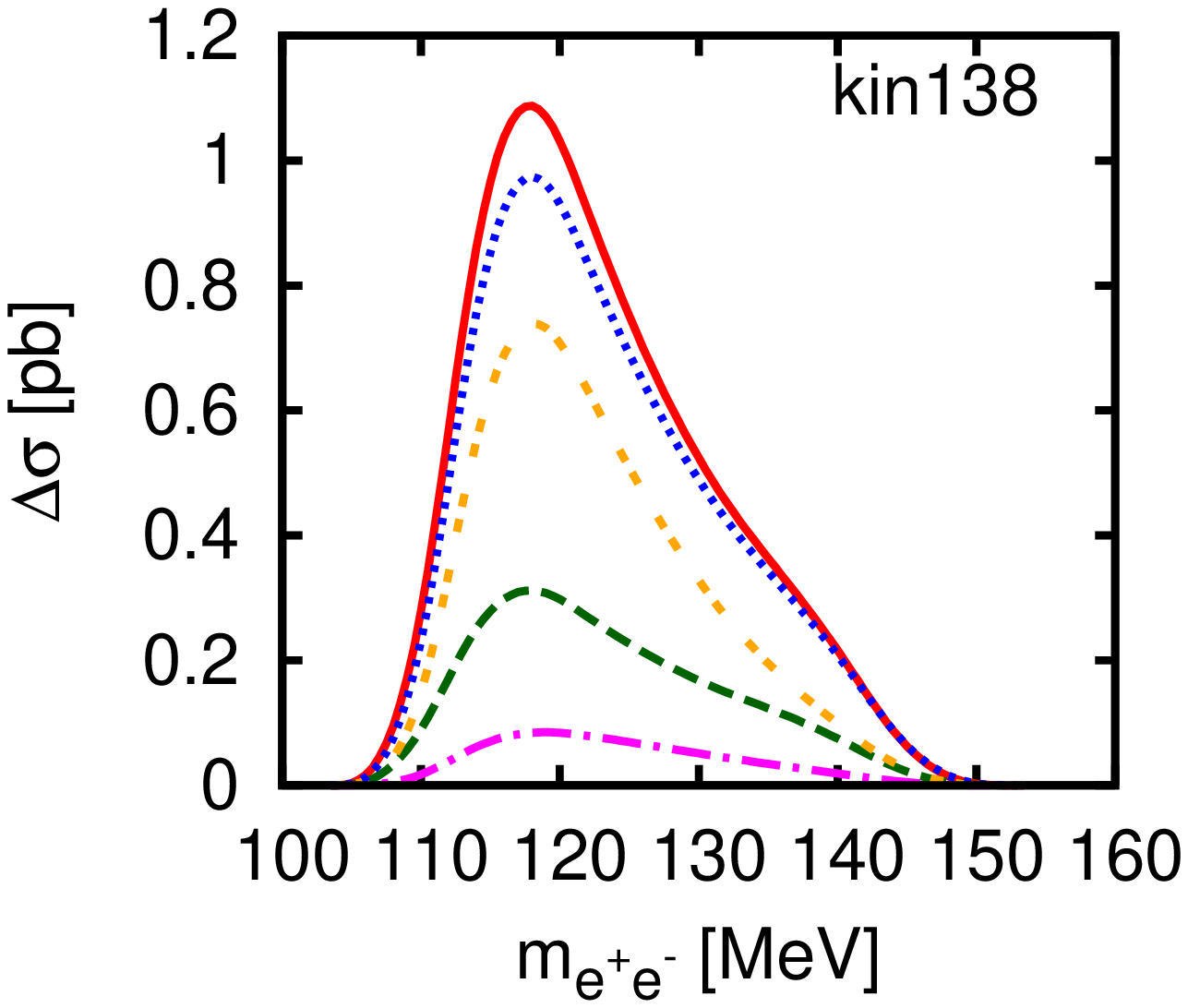}\\
 \includegraphics[angle=-90,width=.32\linewidth]{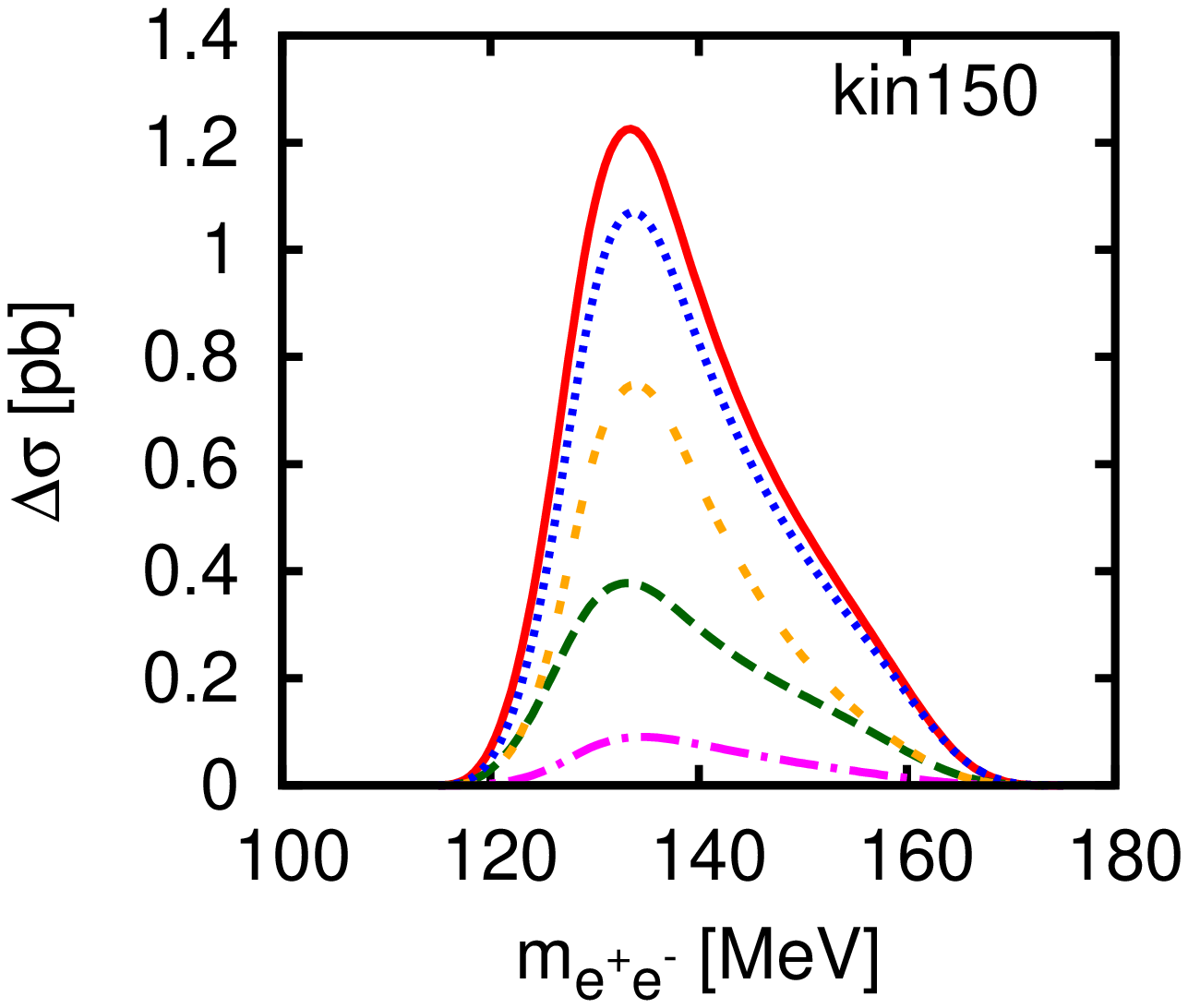}
 \includegraphics[angle=-90,width=.32\linewidth]{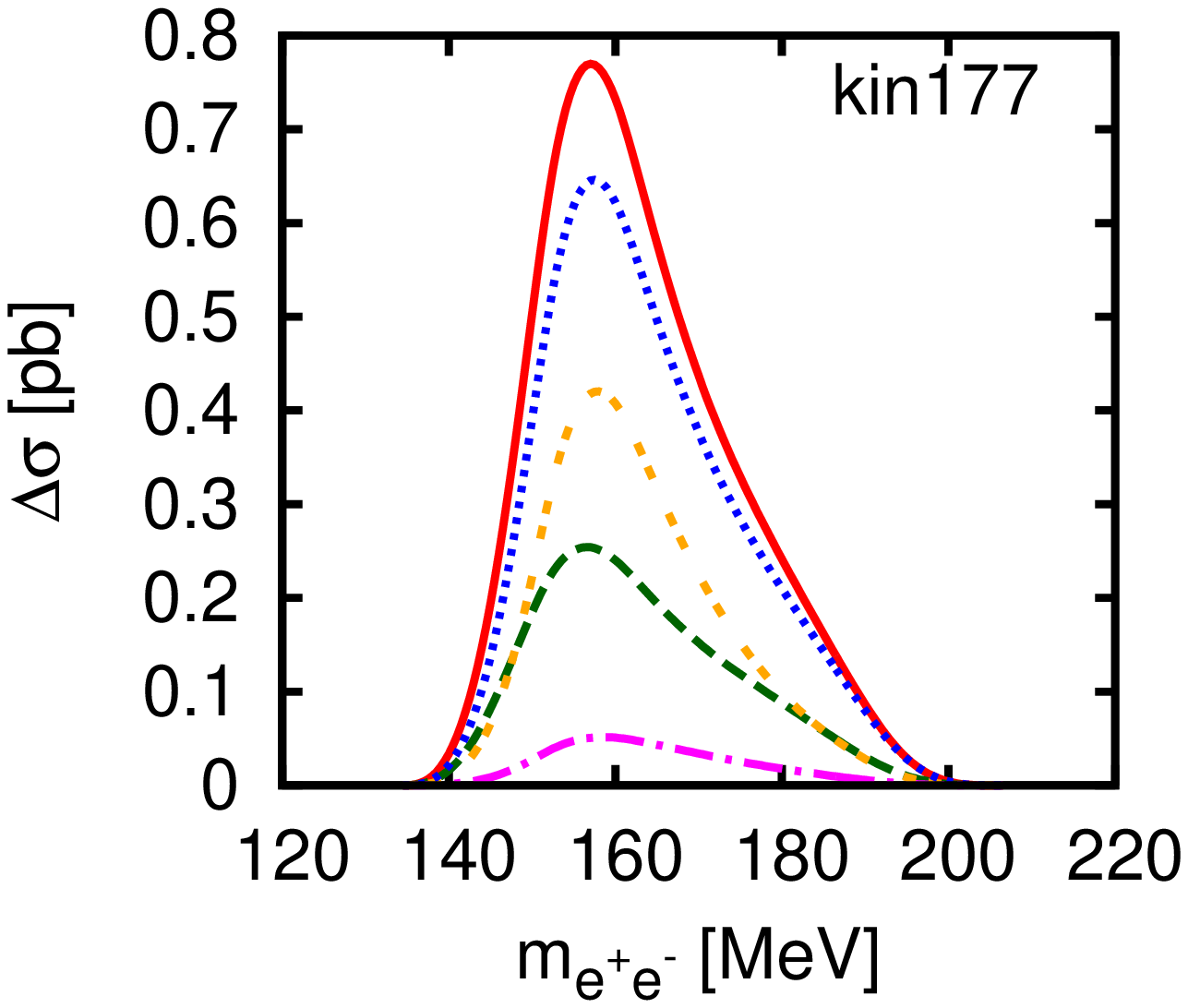}
 \includegraphics[angle=-90,width=.32\linewidth]{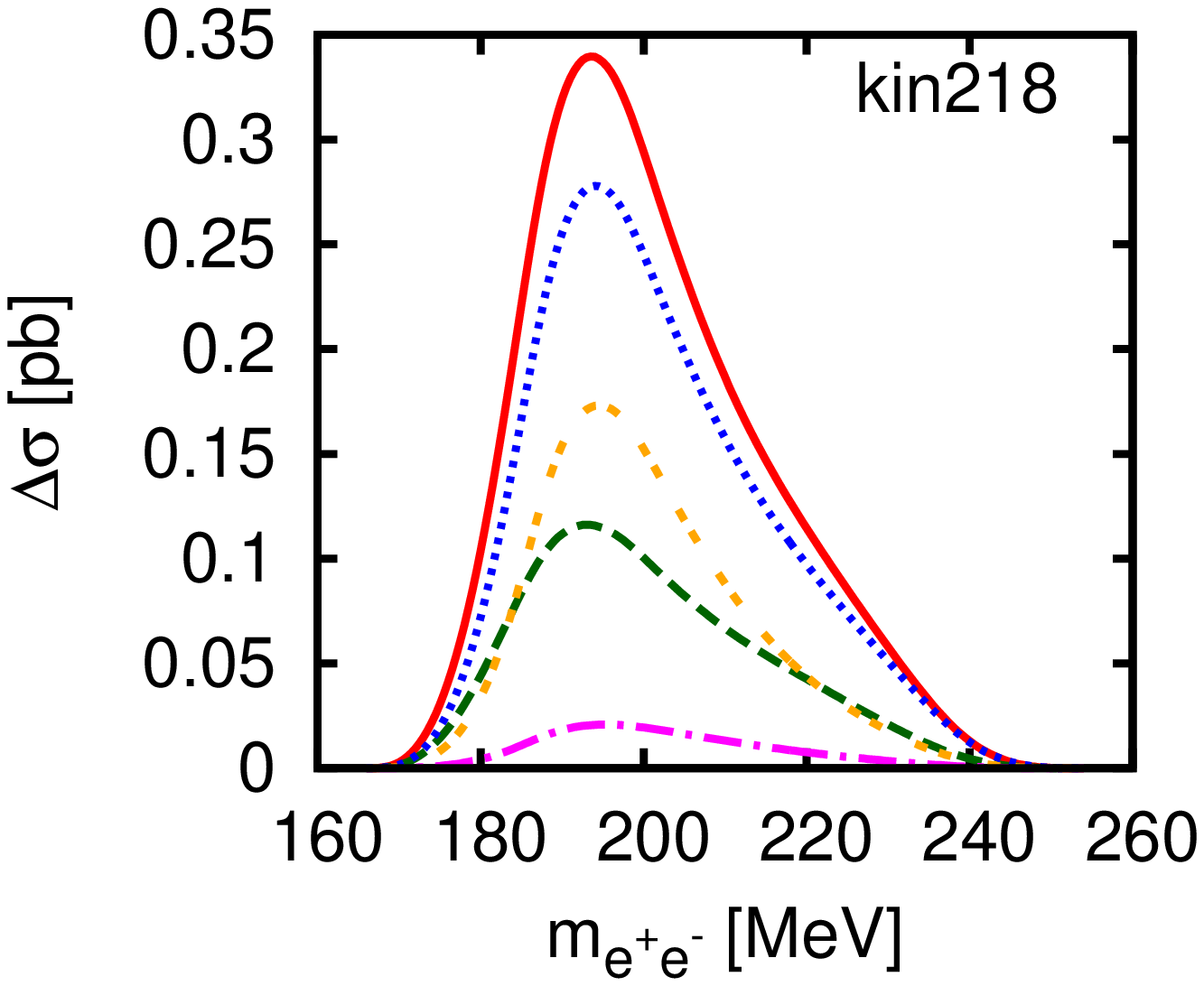}
\caption{Simulation of the invariant mass distributions calculated from the different cross sections for the kinematics probed at MAMI in 2012: background (solid curve), SL + TL exchange term (dotted), SL exchange term (double-dashed), SL + TL direct term (dashed), and TL direct term (dashed-dotted).\label{fig:mami2012_invmass}}
 \end{figure*}
 
\begin{figure}
\includegraphics[angle=-90,width=.5\linewidth]{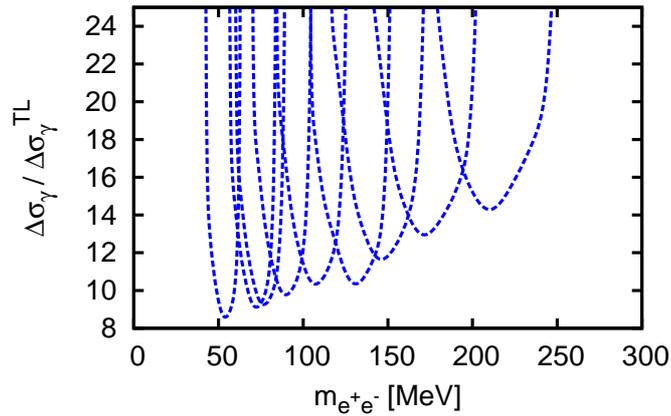}
\caption{Combined plot of our result for the ratios ${\Delta\sigma_\gamma}/{\Delta\sigma_\gamma^\text{TL}}$ of each setting, starting with the lowest beam energy on the left.
\label{fig:mami2012_combined_ratio}}
\end{figure}

The A1 Collaboration started a $\Ap$ search run at MAMI in 2012, probing the kinematics given in Table~\ref{tab:mami2012_kins}, in which no signal of a $\Ap$ was found.
The obtained invariant mass distributions can be seen in Fig.~\ref{fig:mami2012_invmass}.
The invariant mass distributions calculated from the different cross sections are compared: background (solid curve), SL + TL exchange term (dotted), SL exchange term (double-dashed), SL + TL direct term (dashed), and TL direct term (dashed-dotted).
It turns out that the SL exchange process is the largest contribution to the radiative background.
Fig.~\ref{fig:mami2012_invmass} illustrates the dependence of the seperated background contributions on the invariant mass $m_{e^+e^-}$.
At low invariant masses the SL exchange term dominates the cross section.
Although the SL direct and the TL exchange terms become more important for increasing $m_{e^+e^-}$, the SL exchange term remains the largest contribution to the cross section.
The ratio between the TL direct term and the SL exchange term has a similar behavior, retaining nearly the same maximum value in each of the considered settings. Furthermore, Fig.~\ref{fig:mami2012_invmass} shows the importance of the interference parts of the cross section, which are necessary to describe the data correctly.\\
In Fig.~\ref{fig:mami2012_combined_ratio} we present a combined plot of our result for the ratio ${\Delta\sigma_\gamma}/{\Delta\sigma_\gamma^\text{TL}}$ for each setting given in Table~\ref{tab:mami2012_kins}, which is crucial to obtain the exclusion limits on the $\Ap$ mass $m_\Ap$ and its coupling strength $\varepsilon^2$ following Eq.~(\ref{epsextr_eq:epssq}), as function of the invariant mass $m_{e^+e^-}$.
Due to the particular choice of kinematics in that experiment, the ratio ${\Delta\sigma_\gamma}/{\Delta\sigma_\gamma^\text{TL}}$ has a value between $10 - 15$ in the probed mass range.\\
In Fig.~\ref{fig:excl_compilation} our predictions for the exclusion limits on $\varepsilon^2$ for this set of kinematics are indicated by the dashed curve for an assumed integrated luminosity of around $10\,\text{fb}^{-1}$.
\section{Future Searches and Discussion\label{sec:future}}
\begin{figure}
 \includegraphics[width=.3\linewidth,angle=-90]{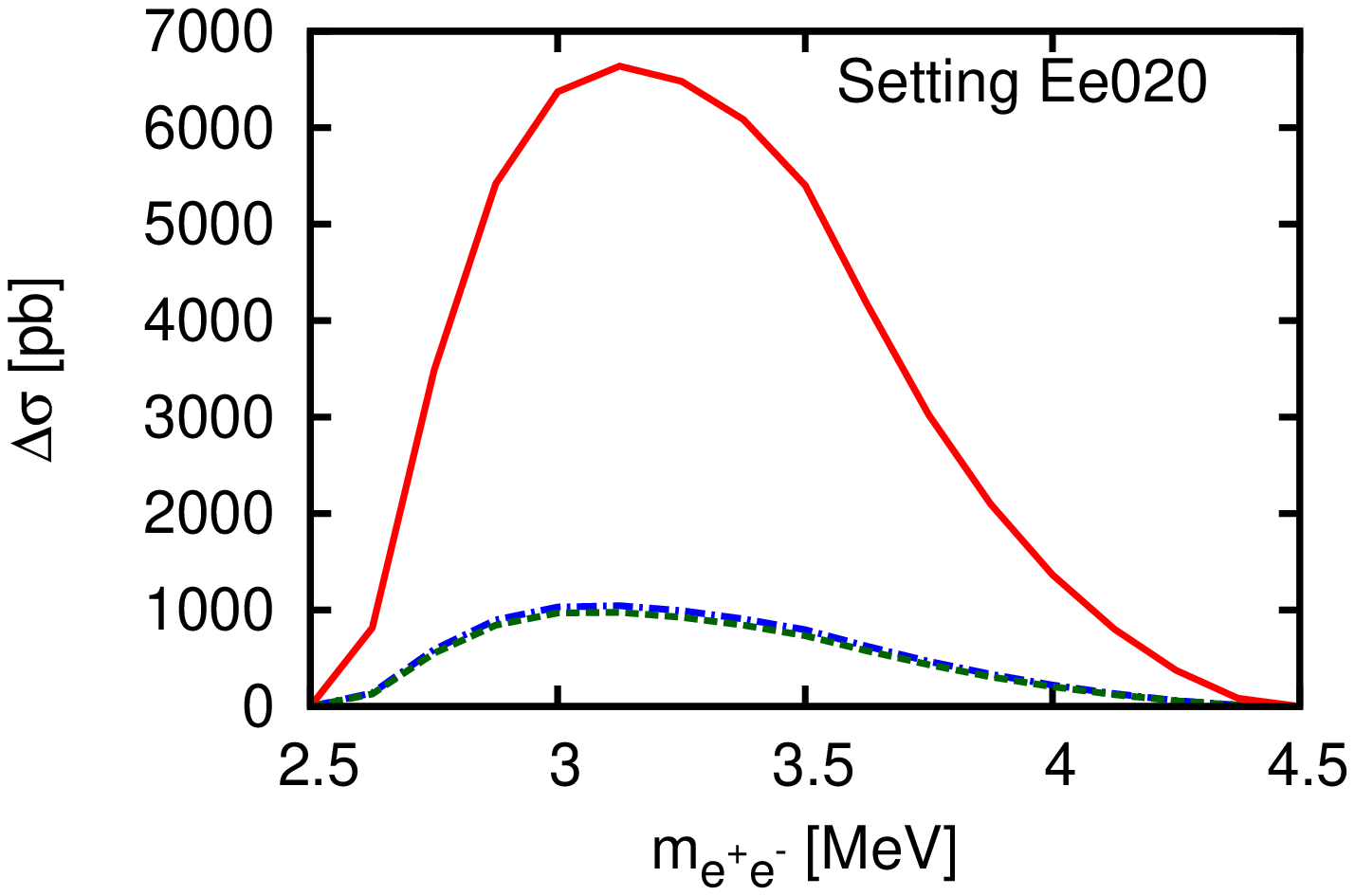}
 \includegraphics[width=.3\linewidth,angle=-90]{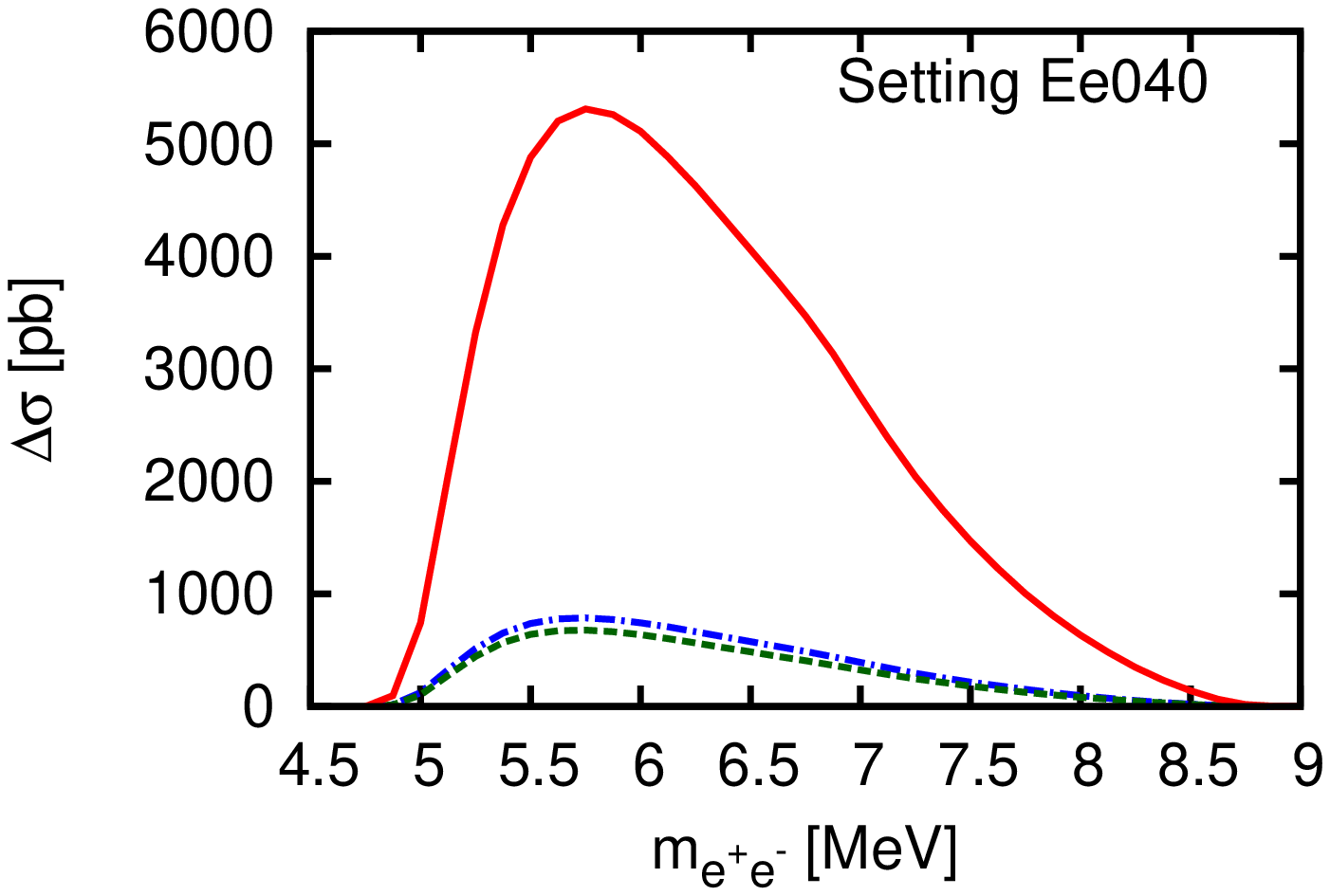}\\
 \includegraphics[width=.3\linewidth,angle=-90]{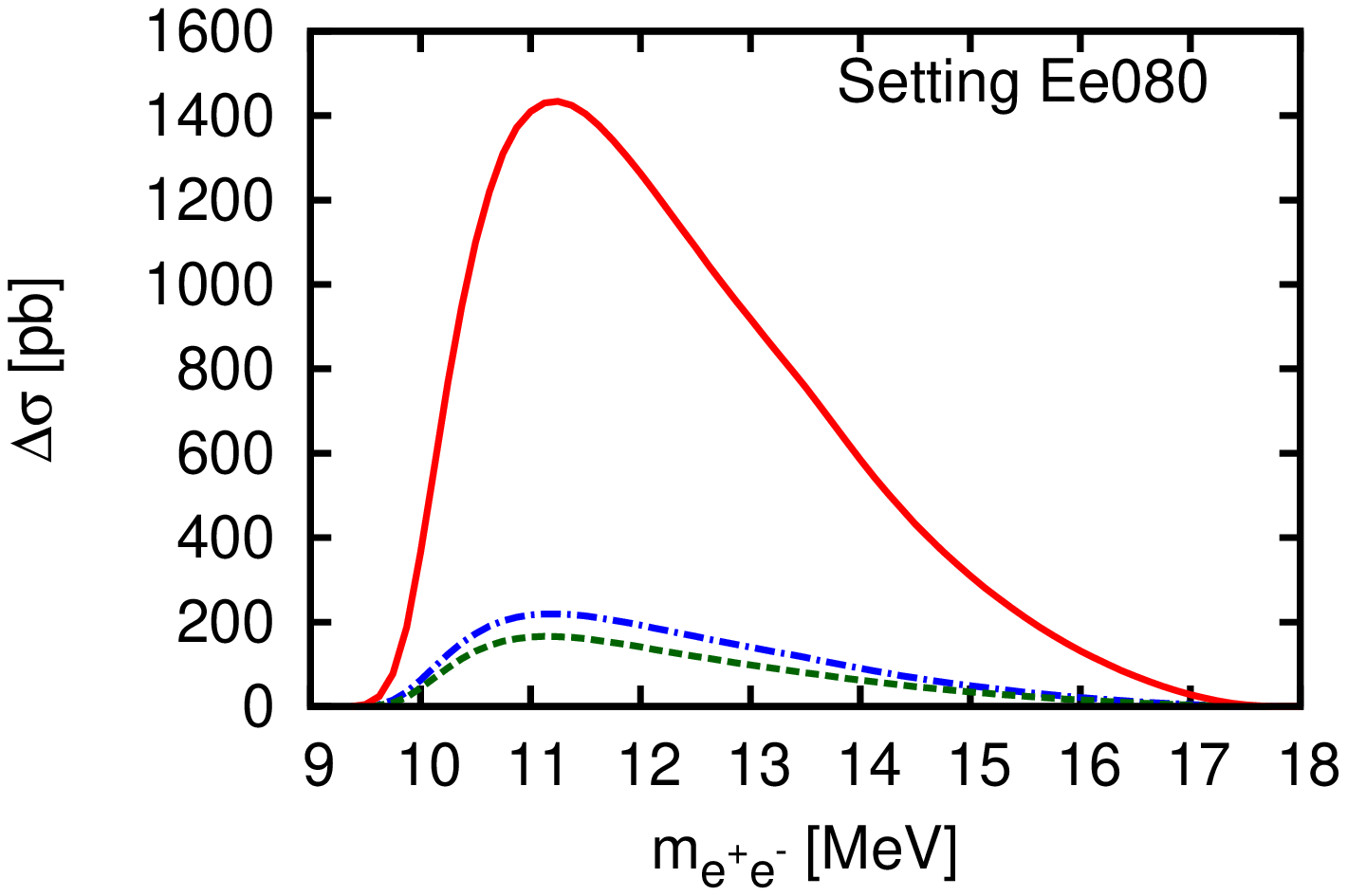}
 \includegraphics[width=.3\linewidth,angle=-90]{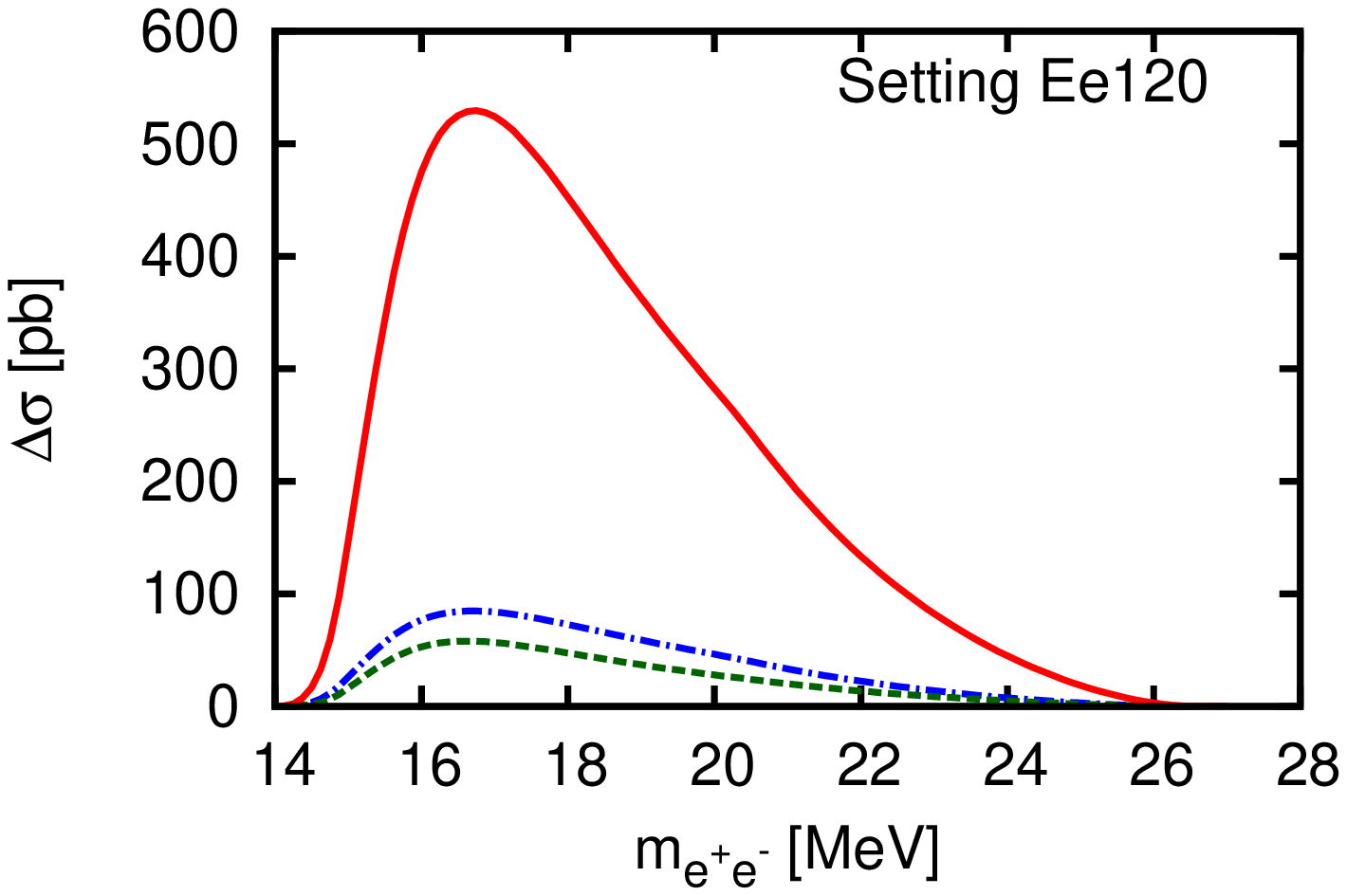}\\
 \includegraphics[width=.3\linewidth,angle=-90]{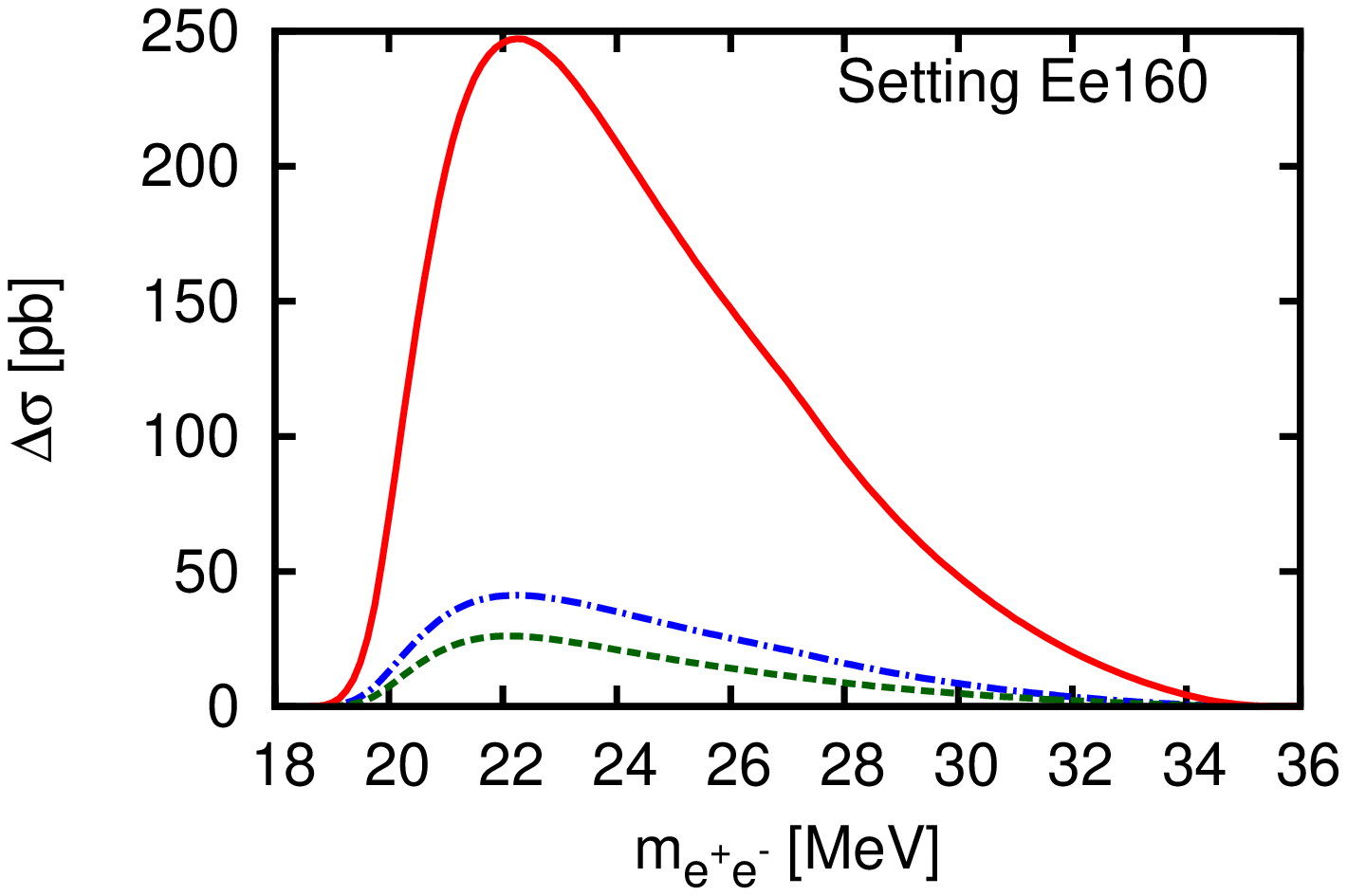}
 \includegraphics[width=.3\linewidth,angle=-90]{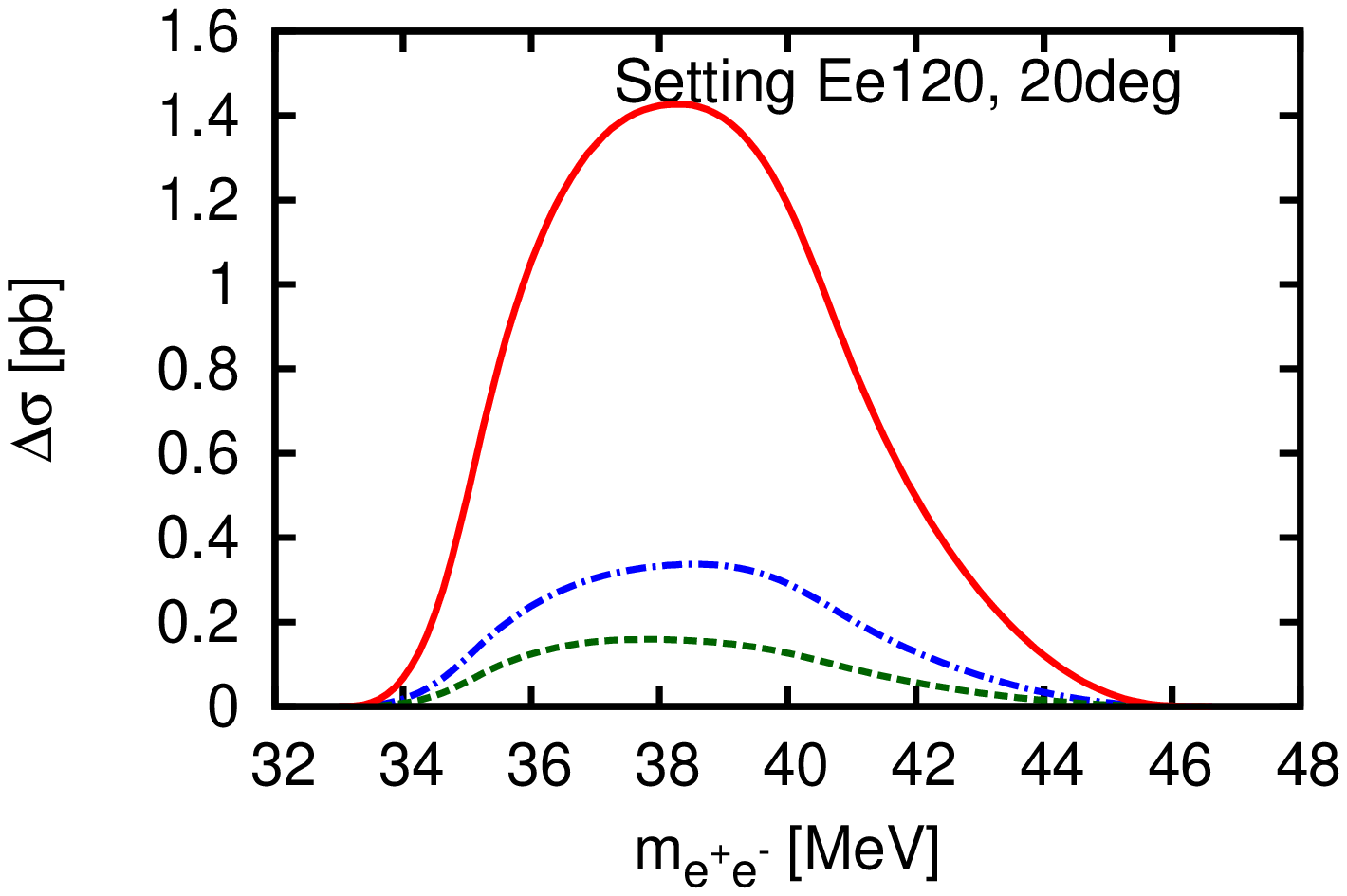}
 \caption{Invariant mass distributions from the feasibility study for the MESA experiment.
Solid curve: SL+TL (direct + exchange term), dashed curve: direct TL, dashed-dotted curve: direct SL+TL.\label{fig:MESA_invmass}}
\end{figure}

\begin{figure}
 \includegraphics[width=.3\linewidth,angle=-90]{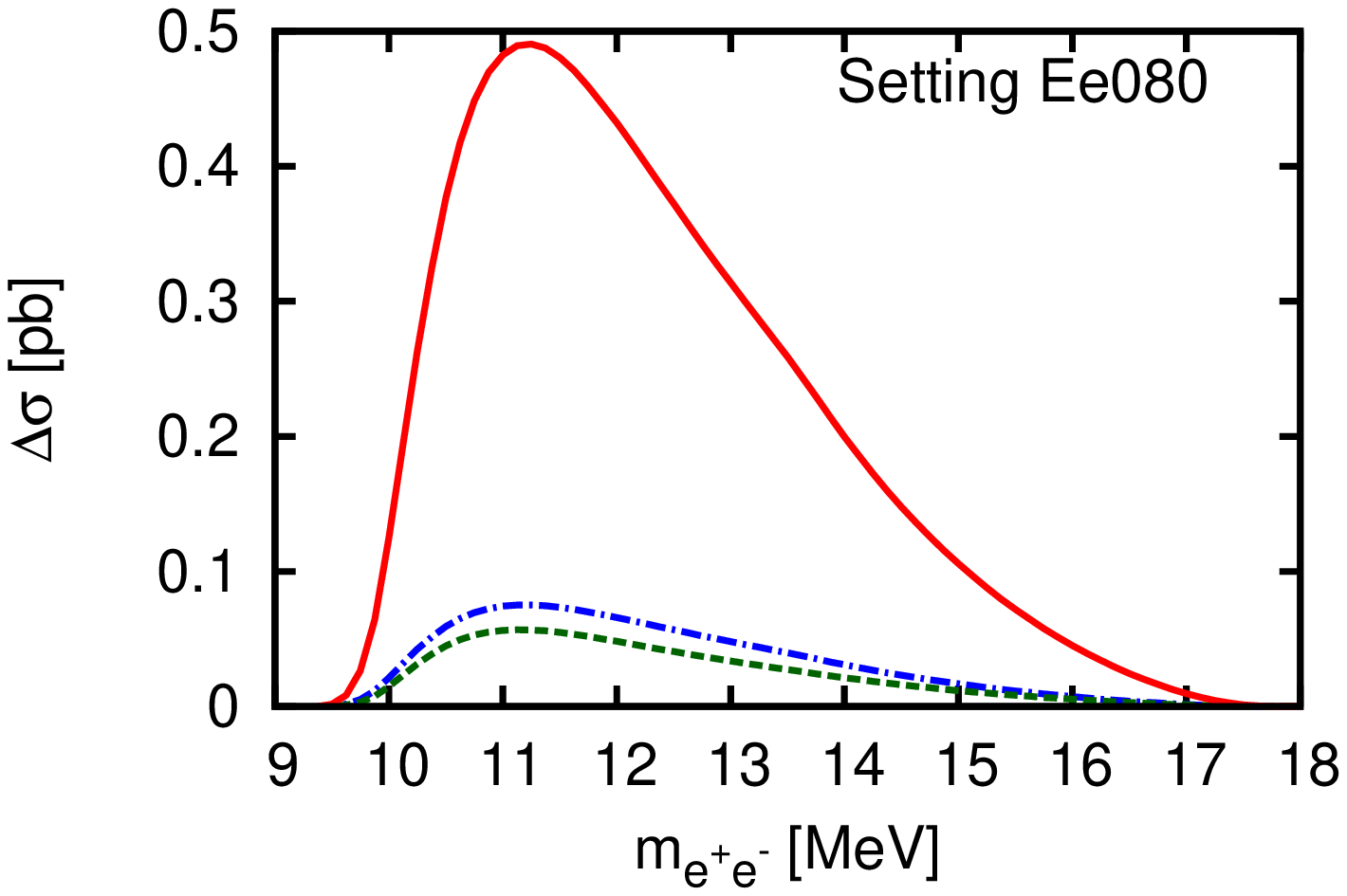}
 \includegraphics[width=.3\linewidth,angle=-90]{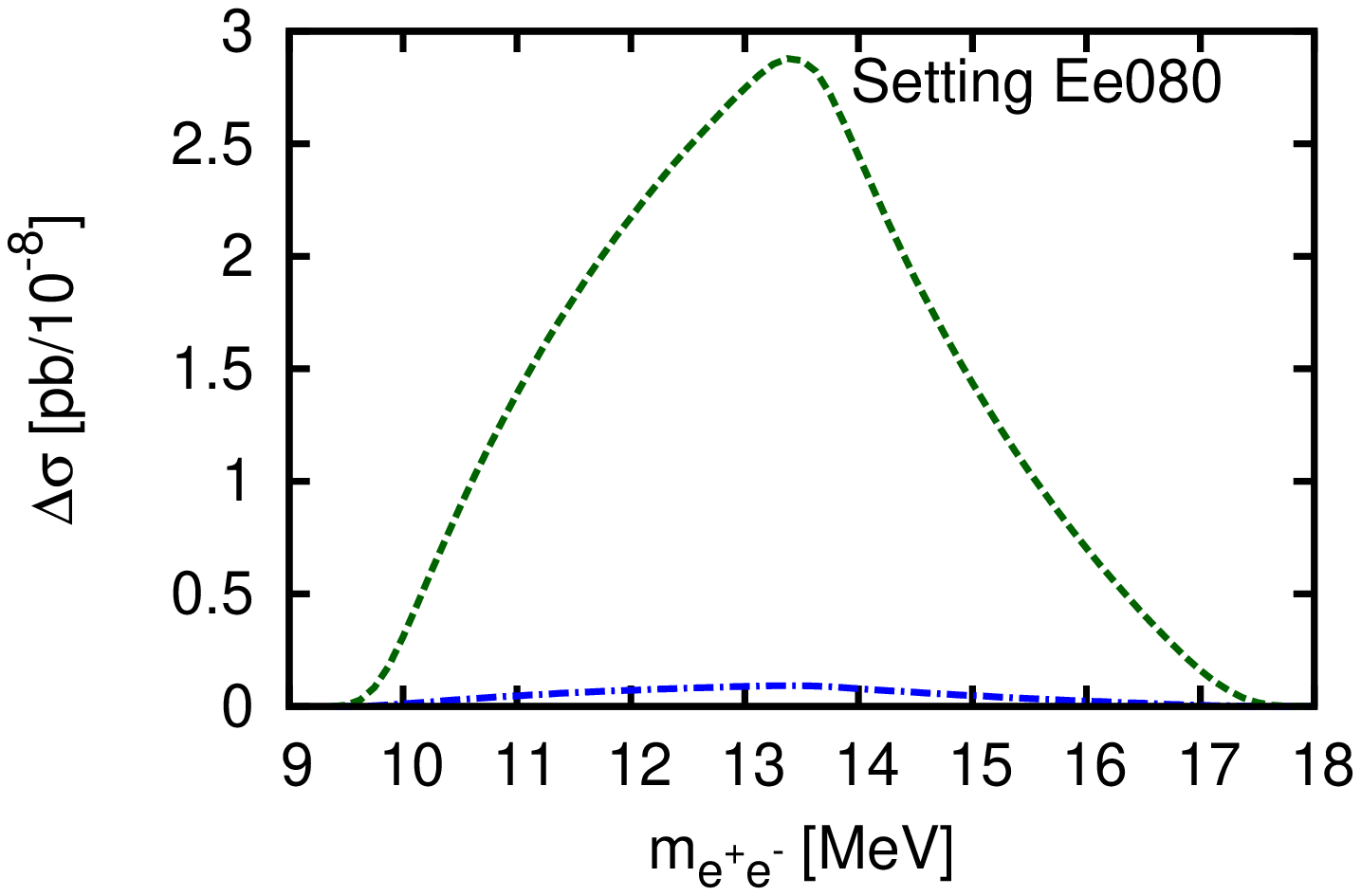}
 \caption{Left panel: Invariant mass distributions from the feasibility study for the MESA experiment for a proton target.
Solid curve: SL+TL (direct + exchange term), dashed curve: direct TL, dashed-dotted curve: direct SL+TL.\\
 Right panel: Isolated VCS cross section.
Dashed curve: exchange term contribution, dashed-dotted curve: direct contribution.\label{fig:MESA_invmass_proton}}
\end{figure}

\begin{figure}
\includegraphics[angle=-90,width=.5\linewidth]{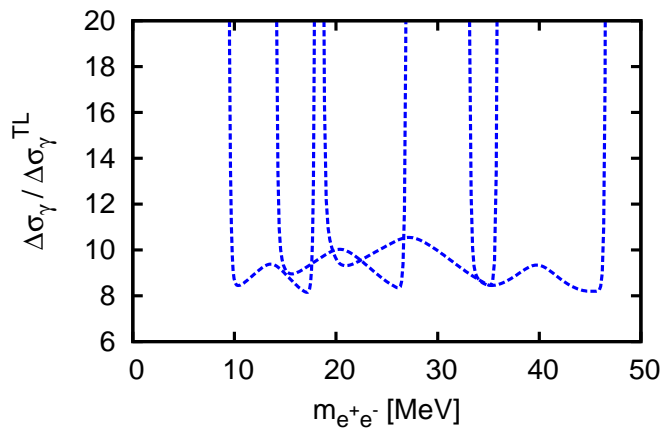}
 \caption{Combined plot of our result for the ratio ${\Delta\sigma_\gamma}/{\Delta\sigma_\gamma^\text{TL}}$ of each setting for the MESA experiment. The settings correspond with the following beam energies and scattering angles (from left to right): $E_0 = 80,\,120,\,160\,\MeV$ with $\phi_\mp=\pm 10^\circ$, $E_0=120\,\MeV$ with $\phi_\mp=\pm 20^\circ$
\label{fig:mesa_combined_ratio}}
\end{figure}

\begin{figure}[t]
 \includegraphics[width=.8\linewidth]{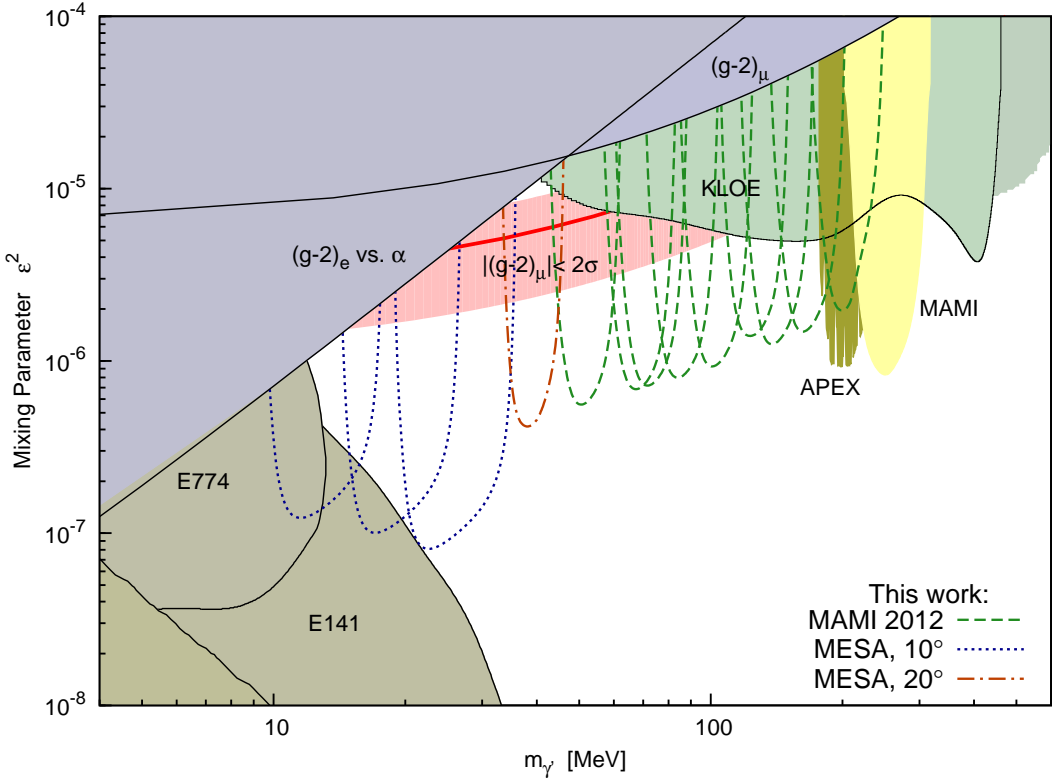}
 \caption{Compilation of existing exclusion limits and our predictions:
 For a better visualization we restrict ourselves to the region currently accessible at fixed-target experiments.
Only existing limits as published in Refs.~\cite{Bjorken:2009mm,Davoudiasl:2012ig,Endo:2012hp,Andreas:2011in,Archilli:2011zc,Babusci:2012cr} are shown, represented by the shaded regions.
We do not show the predictions for other experiments \cite{Essig:2010xa,Freytsis:2009bh,HPS,Kahn:2012br} in this figure which are scheduled to probe the same region of parameter space.
The limits of MAMI  and APEX are those as given in their publications \cite{Merkel:2011ze,Abrahamyan:2011gv}.
The prediction of this work for the exclusion limit expected for the MAMI 2012 experiment discussed in section~\ref{sec:MAMI2012} is depicted by the dashed curves.
The prediction for MESA obtained in section~\ref{sec:future} is indicated by the dotted (dashed-dotted) curve for the setups with a central scattering angle of $10^\circ$ $(20^\circ)$.
\label{fig:excl_compilation}}
\end{figure}

Recently the construction of the Mainz Energy Recovering Accelerator (MESA) has been approved.
MESA is aimed to provide a high intensity electron beam up to beam energies of about $160\,\MeV$ and thus should be ideally suited to probe the $\Ap$ parameter space for low masses.
In this section we perform a feasibility study to carry out this search by using two small spectrometers. We assume, that each of these spectrometers has a horizontal and vertical angular acceptance of $\pm 50\,\text{mrad}$ and a momentum acceptance of $\pm 5\%$.
A possible $\Ap$ experiment at MESA can be performed using a gas target to minimize the multiple scattering in the target material.
Therefore, applying the same program code as in section~\ref{sec:MAMI_results}, we perform our calculations using a Xenon target in order to obtain as large cross sections as possible.
The integration over the invariant mass $m_{e^+e^-}$ is performed for a $0.125\,\MeV$ interval.\\
The results for the obtained invariant mass distributions of this study are shown on Fig.~\ref{fig:MESA_invmass}.
The kinematics were chosen such that the central scattering $\phi$ of the $e^-$ ($e^+$) is $+10^\circ$ ($-10^\circ$) and the central momentum is $\Abs{l}_\pm=0.98\times {E_0}/{2}$ for beam energies $E_0$ of $20$, $40$, $80$, $120$, and $160\,\MeV$.
Furthermore, we have calculated one setting for $E_0= 120\,\MeV$ and $\phi_\mp=\pm 20^\circ$ in order to cover the full so-called $(g-2)_\mu$ welcome band together with the MAMI 2012 settings.
We assume a beam time of about 3 months and a luminosity of $10^{34}\,\text{cm}^{-2}\text{s}^{-1}$.\\
Since the low mass region $m_\Ap\lesssim 10\,\MeV$ in the $(g-2)_\mu$ discrepancy is already excluded by the electron anomalous magnetic moment $(g-2)_e$, the settings for beam energies of $20$ and $40\,\MeV$ will not enter the exclusion limit calculation.
Therefore we do not have to deal with the difficulties in the low mass regime.
From our exact calculation of the signal cross section $\Delta \sigma_{\Ap}$ we find for the considered range of parameters a good agreement with the approximation of the signal cross section given in Ref.~\cite{Bjorken:2009mm}.\\
For comparison we show in Fig.~\ref{fig:MESA_invmass_proton} the acceptance integrated cross section depending on $m_{e^+e^-}$ for a proton target with a beam energy of $E_0 = 80\,\MeV$.
In the left panel the same curves as in Fig.~\ref{fig:MESA_invmass} are plotted.
In the right panel of Fig.~\ref{fig:MESA_invmass_proton} it is demonstrated, that the VCS contribution corresponding with the Feynman diagrams in Fig.~\ref{fig:feyn_ep-epll_vcs} are smaller by more than 6 orders of magnitude in the chosen kinematic setting, and can thus be neglected.
As indicated by the shape of the curves for $\Delta \sigma_{\gamma^\ast,\D+\X}^{SL+TL}$ and $\Delta \sigma_{\gamma^\ast,\D}^{TL}$ in Figs.~\ref{fig:MESA_invmass} and \ref{fig:MESA_invmass_proton}, the ratio of these two quantities is equal and thus the kind of target does not affect the exclusion limit concerning the QED background.\\
Fig.~\ref{fig:mesa_combined_ratio} shows the calculated ratio ${\Delta\sigma_\gamma}/{\Delta\sigma_\gamma^\text{TL}}$ which reaches a value around $8\,-\,10$ for the proposed settings.
The expected exclusion limit on $\varepsilon^2$ as obtained from Eq.~(\ref{epsextr_eq:epssq}), to the invariant mass spectra of Fig.~\ref{fig:MESA_invmass}, is presented on Fig.~\ref{fig:excl_compilation}, where a mass resolution of $0.125\,\MeV$ was assumed.
The dotted (dashed-dotted) curve on Fig.~\ref{fig:excl_compilation} represents the settings with central angle of $10^\circ$ $(20^\circ)$.
At very low masses below $10\,\MeV$ Eq.~(\ref{epsextr_eq:epssq}) does not serve as a good approximation for the exclusion limit anymore, since Eq.~(19) of Ref.~\cite{Bjorken:2009mm} overestimates the $\Ap$ signal cross section by up to $50\%$.\\

A compilation of the existing exclusion limits is presented in Fig.~\ref{fig:excl_compilation}, which shows the region $5\,\MeV \leq m_{\Ap} \leq 600\,\MeV$ and $10^{-8}\leq \varepsilon^2 \leq10^{-4}$ accessible at fixed-target experiments.
Furthermore, existing limits as published in Refs.~\cite{Bjorken:2009mm,Davoudiasl:2012ig,Endo:2012hp,Andreas:2011in,Archilli:2011zc,Babusci:2012cr} are also shown, and are represented by the shaded regions.
Let us mention that other planned experiments \cite{Essig:2010xa,Freytsis:2009bh,HPS,Kahn:2012br} are scheduled to probe the same region of parameter space.
The limits of MAMI  and APEX are those as given in their publications \cite{Merkel:2011ze,Abrahamyan:2011gv}.
Our prediction for the exclusion limit expected in the MAMI 2012 experiment discussed in section~\ref{sec:MAMI2012} is depicted by the dashed curves.
The prediction for MESA obtained in section~\ref{sec:future} is indicated by the dotted (dashed-dotted) curves for the setups with a central scattering angle of $10^\circ$ $(20^\circ)$.\\
Our calculation shows, that the 2012 experiment is well suited to exclude a large region of the parameter space and in particular most of the so-called $(g-2)_\mu$ welcome band, in which the discrepancy between the experimental and theoretical value of the anomalous magnetic moment of the muon $(g-2)_\mu$ could be due to $\Ap$ contribution.\\
We propose an experiment for the MESA accelerator under construction at Mainz.
The investigated kinematic settings will allow for the exclusion of the remaining part of the $(g-2)_\mu$ welcome band that is not probed so far.
\section{Conclusions and Outlook}
In this work we have calculated the cross section which are crucial to describe the existing and planned fixed-target $\Ap$ search experiments.
A comparison of our calculations with a sample of data taken at MAMI has been performed.
After applying the leading order QED radiative corrections for the corresponding elastic electron-hadron scattering process we find, that our calculations and the data sample are in good agreement.
In addition, a calculation of the separated spacelike and timelike virtual photon exchange cross sections, each for the direct and exchange term, has been performed.
This allows us to study the dependence of the background cross section on these contributions.
Furthermore we find, that it is necessary to include the exchange term into the cross section in order to reconcile the data.
The exchange contribution is contributing to the irreducible background.\\
Using the cross sections obtained in our analysis, we are able to provide predictions for the expected exclusion limits for MAMI and MESA.
Following our predictions, the experiments at MAMI and MESA will be able to probe the entire $(g-2)_\mu$ welcome band and in addition, increase the existing limits by one order of magnitude.
\begin{acknowledgments}
This work was supported in part by the Research Centres ``Elementarkr\"afte und Mathematische Grundlagen'' at the Johannes Gutenberg University Mainz,  in part by the Deutsche Forschungsgemeinschaft DFG through the Collaborative Research Center ``The Low-Energy Frontier of the Standard Model'' (SFB 1044), and the federal state of Rhineland-Palatinate.
TB likes to thank Bj\"orn Walk for helpful discussions on GPU programming. Furthermore, we thank Achim Denig for useful discussions.
\end{acknowledgments}

\appendix
\section{Detailed Cross Section Calculations for \texorpdfstring{$\Ap$}{A'} Search experiments}\label{app_xs-derivation}
Starting from Eq.~(\ref{eq_dcs-gen}) one finds by inserting $1 = \int d^4 q^\prime\, \delta^{(4)}\!\left(q^\prime - l_+ - l_- \right)$
\begin{align}
d \sigma
   &=\frac{1}{4\sqrt{(k \cdot p)^2-m^2M^2}} \frac{d^3\vecp{k}}{(2\pi)^3\,2\, E_e^\prime}\, \frac{d^3\vecp{p}}{(2\pi)^3\,2\,E_p^\prime}\, \frac{d^3\vecp{q}}{(2\pi)^3\,2\,q^{\prime 0}}\nonumber\\
   &\quad (2\pi)^4\delta^{(4)}\left(k+p-k^\prime-p^\prime-q^\prime\right)\cdot  \frac{d^3\vec{l_-}}{(2\pi)^3\,2\,E_-}\,   \frac{d^3\vec{l_+}}{(2\pi)^3\,2\,E_+}\nonumber\\
 &\quad {}\cdot \underbrace{\frac{q^{\prime 0}\,d q^{\prime 0}}{2\pi}}_{d q^{\prime\,2}/(2\pi)}\,(2\pi)^4\, \delta^{(4)}\!\left(q^\prime - l_+ - l_- \right)\overline{\left|\mathcal{M}\right|^2}.\nonumber
\end{align}
The $\delta$-functions constrain the three-momenta
\[
\vec{q}=\vec{l_-} + \vec{l_+}\quad \text{and} \quad \vecp{p}=\vec{k} - \vecp{k} - \vecp{q},
\]
which leads to
\begin{align}
 d \sigma&= \frac{1}{128\,\Abs{k}\,M} \frac{1}{\left(2\pi\right)^8}\frac{\Absp{k}^2\Abs{l_+}^2\Abs{l_-}^2}{E_{p^\prime}E_{k^\prime}E_{A^\prime}E_{l_+}E_{l_-}}  d\Absp{k}\,d\Omega_{e^\prime}\, d\Abs{l_+}\,d\Omega_+\,d\Abs{l_-}\,d\Omega_-\,dq^{\prime\,2}\nonumber\\
&\quad {} \delta\underbrace{\left(E_0+M-E_{e^\prime}-E_{p^\prime}-q^{\prime 0}\right)}_{=:\delta_1}\,\delta \underbrace{\left(q^{\prime 0} - E_+ - E_-\right)}_{=:\delta_2}\overline{\left|\mathcal{M}\right|^2}
\label{epepll:dcs8-01}.
\end{align}
The remaining two delta functions can be used to express the energies associated with $k^\prime$ and $l_-$, by which integration over their three-momentum absolute values is performed.
Therefore expressions for $\Absp{k}$ and $\Abs{l_-}$ in terms of the remaining quantities have to be found, using
\begin{align}
 &q^{\prime 2} = 2 m_l^2 + 2 E_+ E_- - 2 \Abs{l_-} \vec{l_+}\cdot \hat{l}_-\nonumber\\
&\Leftrightarrow 0 = \underbrace{\left( -\frac{q^{\prime 2}}{2} +m_l^2\right)}_{=:A} + E_+ E_- - \Abs{l_-}
\underbrace{\vec{l_+}\cdot \hat{l}_-}_{=:B}\label{eq:epepll_norm_l-_1}.
\end{align}
This equation can be rewritten as a quadratic equation for $\Abs{l_-}$ which can be easily solved.
After adding $(B \Abs{l_-} - A)$ on both sides of Eq.~(\ref{eq:epepll_norm_l-_1}), squaring the result and using $E_-^2=\Abs{l_-}^2+m_l^2$ one finds the two solutions
\begin{align}
\Abs{l_-}_{1,2} = \frac{A B}{B^2 - E_+^2} \pm \sqrt{ \frac{(A E_+)^2 + (E_+ m_l B)^2 - (E_+^2m_l)^2}{(B^2-E_+^2)^2} }.\label{eq:epepll_norm_l-2}
\end{align}
The determination of the physical solution can be done by considering the particles as massless.
Now the calculation simplifies to
\begin{align}
 & q^{\prime 2} = \underbrace{l_+^2 + l_-^2}_{= 0} + 2 \Abs{l_-} \Abs{l_+} (1 - \hat{l}_+ \cdot \hat{l}_-) \nonumber\\
&\Leftrightarrow \Abs{l_-} = \frac{q^{\prime 2}}{2 \Abs{l_+} (1 - \hat{l}_+ \cdot \hat{l}_-) }.
\label{eq:epepll_norm_l-_massless_final}
\end{align}
Comparing Eqs.~(\ref{eq:epepll_norm_l-2}) and (\ref{eq:epepll_norm_l-_massless_final}), one finds that the solution with ``$+$'' corresponds to the physical allowed case.
Thus it is
\begin{equation}
 \Abs{l_-} = \frac{A B}{B^2 - E_+^2} + \sqrt{ \frac{(A E_+)^2 + (E_+ m_l B)^2 - (E_+^2m_l)^2}{(B^2-E_+^2)^2} },\label{eq:epepll_norm_l-final}
\end{equation}
with $A =- {q^{\prime 2}}/{2} +m^2$ and $B=\vec{l_+}\cdot \hat{l}_-$.\\
The calculation of $\Absp{k}$ is done in a similar way.
Since it is not necessary that the four-vectors $l_+$ and $l_-$ appear explicitely in the following, instead their sum $q^{\prime 2} = (l_+ + l_-)^2$ is used where $\Abs{l_-}$ is symbolic for the result of Eq.~(\ref{eq:epepll_norm_l-final}).
Again starting from four-momentum conservation one finds
\begin{align}
\Leftrightarrow 0 =\underbrace{(p+k-q^\prime)^2+m^2-M^2 }_{=:D} - \underbrace{2 (E_0 + M - q^{\prime 0})}_{=:F} E_{e^\prime} + \underbrace{2 (\vec{k}-\vecp{q})\cdot \hat{k}^\prime}_{=:G} \Absp{k}.\nonumber
\end{align}
An analogous calculation as for $\Abs{l_-}$ then leads to
\begin{equation}
 \Absp{k} = -\frac{D G}{G^2 - F^2} + \sqrt{ \frac{(m F G)^2 + (D F)^2 - ( m F^2 )^2}{(G^2-F^2)^2} }.\label{eq:epepll_norm_kp_final}
\end{equation}
Thus one has
\begin{align}
 \frac{\partial \delta_1}{\partial \Absp{k}} &= \frac{\partial}{\partial \Absp{k}} \left(E_0+M-E_{e^\prime}-E_{p^\prime}-q^{\prime 0} \right)\nonumber\\
&= -\frac{\Absp{k}}{E_{k^\prime}}-\frac{\Absp{k}-\hat{k}^\prime\cdot \left(\vec{k}- \vecp{q}\right)}{E_{p^\prime}}
\label{eq_epepll:ddelta1}
\end{align}
and
\begin{align}
 \frac{\partial \delta_2}{\partial \Abs{l_-}} &= \frac{\partial}{\partial \Abs{l_-}} \left( q^{\prime 0} - E_+ - E_-\right)\nonumber\\
&= -\frac{\Abs{l_-}}{E_-}+\frac{\Abs{l_-} + \vec{l_+}\cdot \hat{l}_-}{ q^{\prime 0} } \label{eq_epepll:ddelta2}.
\end{align}

For the experiments performed at MAMI the detector quantities are given in table \ref{tab:MAMI_acceptance}.
The horizontal and vertical acceptances are given in a Cartesian reference frame.
It is convenient to calculate the cross section directly in the lab frame.
The lab frame three-momenta of the detected particles depending on these quantities are parametrized by
\begin{equation*}
 \vec{l_\pm} = \frac{\Abs{l_\pm}}{\sqrt{1+\tan^2{\delta\theta} + \tan^2{\delta\phi}}} \left(
 \begin{array}{c}
  \tan{\delta\phi}\, \cos{\phi_0} + \sin{\phi_0}\\
  \tan{\delta\theta}\\
  \cos{\phi_0} - \tan{\delta\phi}\,  \sin{\phi_0} 
 \end{array}
\right)
 ,
\end{equation*}
where $\phi_0$ is the central horizontal angle of the detector, $\delta\phi$ is the deviation from the horizontal scattering angle and $\delta\theta$ is the deviation from the vertical out-of-plane angle.
Note that the vertical central angle of the detectors is $0^\circ$.
Integrating over the angles $\delta\phi$ and $\delta\theta$ within the limits of the experimental acceptances then leads to the cross section $\Delta \sigma$.
To account for this geometry the cross section has to be multiplied by a Jacobian
\[
J(\delta \phi,\, \delta \theta) = \left| \frac{1}{\cos^2{\delta \phi}\,\cos^2{\delta \theta} \left({1+\tan^2{\delta\theta} + \tan^2{\delta\phi}} \right)^{3/2}} \right| .
\]
The cross section then reads as
\begin{align}
&\frac{d \sigma}{d\Abs{l_+} \,d\Omega_+\,d\Omega_-\,d\Omega_{e^\prime}\,dq^{\prime\,2}} \nonumber\\
&=\frac{1}{128\,\Abs{k}\,M} \frac{1}{\left(2\pi\right)^8}\frac{\Absp{k}^2\Abs{l_+}^2\Abs{l_-}^2} {E_{p^\prime}E_{k^\prime}E_{A^\prime}E_{+}E_{-}} J(\delta \phi_-,\, \delta \theta_-) J(\delta \phi_+,\, \delta \theta_+)
 \left(\left| \frac{\partial \delta_1}{\partial \Absp{k}}\right|\left| \frac{\partial \delta_2}{\partial \Abs{l_-}} \right| \right)^{-1}\overline{\left|\mathcal{M}\right|^2}.
\label{eq_epepll:dxs8-2}
\end{align}

\end{document}